\documentclass[12pt,a4paper]{article}
\pdfoutput=1

\usepackage{amsmath,amssymb}
\usepackage[bookmarks=true]{hyperref}
\usepackage[nosort]{cite}
\usepackage[bulletsep]{collect}

\ifx\hypersetup\sadfkjashdfkxja\else
\hypersetup{pdftitle={The Classical r-matrix of AdS/CFT and its Lie Bialgebra Structure}}
\hypersetup{pdfauthor={Niklas Beisert, Fabian Spill}}
\hypersetup{pdfsubject={Mathematical Physics}}
\hypersetup{pdfkeywords={AdS/CFT, Lie Bialgebra, Classical r-matrix, u(2|2)}}
\hypersetup{bookmarksnumbered=true}
\hypersetup{pdfstartview=FitH}
\hypersetup{pdfpagemode=None}
\hypersetup{colorlinks=false}
\hypersetup{citebordercolor={.5 1 .5}}
\hypersetup{urlbordercolor={.5 1 1}}
\hypersetup{linkbordercolor={1 .7 .7}}
\fi

\makeatletter
\let\old@startsection=\@startsection
\renewcommand{\@startsection}[6]{\old@startsection{#1}{#2}{#3}{#4}{#5}{#6\mathversion{bold}}}
\makeatother

\allowdisplaybreaks[3]

\makeatletter \@addtoreset{equation}{section} \makeatother

\makeatletter
\let\old@makecaption=\@makecaption
\def\@makecaption{\small\old@makecaption}
\makeatother

\let\oldPhi=\Phi
\let\oldPsi=\Psi
\let\oldGamma=\Gamma
\let\oldDelta=\Delta
\let\oldSigma=\Sigma
\let\oldLambda=\Lambda
\let\oldTheta=\Theta
\let\oldPi=\Pi
\renewcommand{\Phi}{\mathnormal{\oldPhi}}
\renewcommand{\Psi}{\mathnormal{\oldPsi}}
\renewcommand{\Gamma}{\mathnormal{\oldGamma}}
\renewcommand{\Sigma}{\mathnormal{\oldSigma}}
\renewcommand{\Delta}{\mathnormal{\oldDelta}}
\renewcommand{\Theta}{\mathnormal{\oldTheta}}
\renewcommand{\Lambda}{\mathnormal{\oldLambda}}
\renewcommand{\Pi}{\mathnormal{\oldPi}}

\newcommand{\gen}[1]{\mathfrak{#1}}

\newcommand{\smat}{\mathcal{S}}
\newcommand{\rmat}{\mathcal{R}}

\newcommand{\crmat}{r}

\newcommand{\perm}{\mathcal{P}}
\newcommand{\eip}{\mathcal{U}}

\newcommand{\superN}{\mathcal{N}}
\newcommand{\copro}{\oldDelta}
\newcommand{\coprocl}{\delta}
\newcommand{\coproop}{\widetilde{\oldDelta}}

\newcommand{\psu}{\alg{psu}(2|2)}
\newcommand{\psucentral}{\alg{psu}(2|2)\ltimes\mathbb{R}^3}
\newcommand{\cybe}[2]{[[#1,#2]]}

\newcommand{\order}[1]{\mathcal{O}(#1)}

\newcommand{\Integers}{\mathbb{Z}}
\newcommand{\Complex}{\mathbb{C}}
\newcommand{\Reals}{\mathbb{R}}

\ifx\genfrac\sdflkaj

\else

\fi
\newcommand{\sfrac}[2]{{\textstyle\frac{#1}{#2}}}
\newcommand{\half}{\sfrac{1}{2}}
\newcommand{\ihalf}{\sfrac{i}{2}}
\newcommand{\quarter}{\sfrac{1}{4}}

\newcommand{\indup}[1]{_{\mathrm{#1}}}

\newcommand{\lrbrk}[1]{\left(#1\right)}
\newcommand{\bigbrk}[1]{\bigl(#1\bigr)}

\newcommand{\bigcomm}[2]{\big[#1,#2\big]}
\newcommand{\comm}[2]{[#1,#2]}

\newcommand{\acomm}[2]{\{#1,#2\}}
\newcommand{\bigacomm}[2]{\big\{#1,#2\big\}}

\newcommand{\set}[1]{\{#1\}}
\newcommand{\state}[1]{\mathopen{|}#1\mathclose{\rangle}}

\newcommand{\xp}[1]{x^{+}_{#1}}
\newcommand{\xm}[1]{x^{-}_{#1}}
\newcommand{\xpm}[1]{x^{\pm}_{#1}}

\newcommand{\gat}[1]{\tilde{\gamma}_{#1}}
\newcommand{\tu}[1]{\tilde{u}_{#1}}

\newcommand{\alg}[1]{\mathfrak{#1}}
\newcommand{\grp}[1]{\mathrm{#1}}

\newcommand{\nn}{\nonumber}
\newcommand{\nln}{\nonumber\\}
\newcommand{\nl}[1][0pt]{\nonumber\\[#1]&\hspace{-4\arraycolsep}&\mathord{}}

\newcommand{\earel}[1]{\mathrel{}&\hspace{-2\arraycolsep}#1\hspace{-2\arraycolsep}&\mathrel{}}
\newcommand{\eq}{\earel{=}}
\newcommand{\beq}{\begin{equation}}
\newcommand{\eeq}{\end{equation}}

\def\[{\begin{equation}}
\def\]{\end{equation}}
\def\<{\begin{eqnarray}}
\def\>{\end{eqnarray}}

\makeatletter
\def\mr@ignsp#1 {\ifx\:#1\@empty\else #1\expandafter\mr@ignsp\fi}%
\newcommand{\multiref}[1]{\begingroup%\let\protect\string%
\xdef\mr@no@sparg{\expandafter\mr@ignsp#1 \: }%
\def\mr@comma{}%
\@for\mr@refs:=\mr@no@sparg\do{\mr@comma\def\mr@comma{,}\ref{\mr@refs}}%
\endgroup}
\makeatother

\newcommand{\hypref}[2]{\ifx\href\asklfhas #2\else\href{#1}{#2}\fi}

\newcommand{\secref}[1]{Sec.~\multiref{#1}}

\newcommand{\tabref}[1]{Tab.~\multiref{#1}}

\renewcommand{\eqref}[1]{(\multiref{#1})}

\ifx\href\asklfhas\newcommand{\href}[2]{#2}\fi
\newcommand{\arxivlink}[1]{\href{http://arxiv.org/abs/#1}{arxiv:#1}}

\begin{document}

\pagenumbering{roman}
\thispagestyle{empty}
\begin{flushright}\footnotesize
\texttt{\arxivlink{0708.1762}}\\
\texttt{AEI-2007-116}\\
\texttt{HU-EP-07/31}
\end{flushright}
\vspace{1cm}

\begin{center}%
{\Large\textbf{\mathversion{bold}%
The Classical r-matrix of AdS/CFT\\
and its Lie Bialgebra Structure
}\par} \vspace{1cm}%

\textsc{Niklas Beisert$^a$ and Fabian Spill$^b$}\vspace{5mm}%

\textit{$^a$ Max-Planck-Institut f\"ur Gravitationsphysik\\%
Albert-Einstein-Institut\\%
Am M\"uhlenberg 1, 14476 Potsdam, Germany}\vspace{3mm}%

\textit{$^b$ Humboldt-Universit\"at zu Berlin, Institut f\"ur Physik,\\%
Newtonstra\ss{}e 15, D-12489 Berlin, Germany}\vspace{3mm}%

\texttt{nbeisert@aei.mpg.de}\\
\texttt{spill@physik.hu-berlin.de}%
\par\vspace{1cm}

\textbf{Abstract}\vspace{7mm}

\begin{minipage}{12.7cm}
In this paper we investigate the algebraic structure of AdS/CFT in the strong-coupling limit.
We propose an expression for the classical r-matrix with (deformed) $\alg{u}(2|2)$ symmetry,
which leads to a quasi-triangular Lie bialgebra
as the underlying symmetry algebra.
On the fundamental representation our r-matrix coincides with the classical limit
of the quantum R-matrix.
\end{minipage}

\end{center}

%%%%%%%%%%%%%%%%%%%%%%%%%%%%%%%%%%%%%%%%%%%%%%%%%%%%%%%%%%%%%%%%%%%%%%%%%%%%%%%%
\newpage
\pagenumbering{arabic}
\setcounter{page}{1}
\renewcommand{\thefootnote}{\arabic{footnote}}
\setcounter{footnote}{0}

%\tableofcontents

%%%%%%%%%%%%%%%%%%%%%%%%%%%%%%%%%%%%%%%%%%%%%%%%%%%%%%%%%%%%%%%%%%%%%%%%%%%%%%%%
%%%%%%%%%%%%%%%%%%%%%%%%%%%%%%%%%%%%%%%%%%%%%%%%%%%%%%%%%%%%%%%%%%%%%%%%%%%%%%%%
\section{Introduction and Overview}

In the last five years since the discovery of integrable structures
in the AdS/CFT correspondence \cite{Minahan:2002ve,Beisert:2003tq,Bena:2003wd,Beisert:2003yb}
much progress has been done towards a complete solution of the two sides of the correspondence,
superstring theory on $AdS_5\times S^5$ and $\mathcal N = 4$ SYM theory,
in the large-$N\indup{c}$ limit.
Indeed, assuming integrability,
long-range Bethe Ans\"atze which fully describe the asymptotic spectrum
of long operators or string states with large spins have been proposed \cite{Beisert:2005fw}.
These Bethe Ans\"atze use the factorised S-matrix
\cite{Staudacher:2004tk} which describes the scattering of elementary excitations.
For a generic integrable model, where our excitations, or magnons in the spin chain picture,
carry momentum or rapidity $u$ it is not sufficient to work
only with an ordinary semi-simple Lie algebra $\alg{g}$,
instead we usually work with the loop algebra $\alg{g}[u,u^{-1}]$,
its affinisation $\hat{g}$
or with related deformations of these structures to account for the spectral parameter.
These deformations include (double) Yangians $\grp{DY}(\alg{g})$
and quantum affine algebras $\grp{U}_q(\hat{\alg{g}})$,
which lead to rational and trigonometric S-matrices on evaluation representations,
where the loop variable $u$ simply takes the value of some complex number
and has the physical interpretation as the magnon rapidity.
We can get these S-matrices on evaluation representations
by solving the invariance equation $\comm{\copro(\gen{J}^A)}{\smat} = 0$ for a minimal set
of generators $\gen{J}^A$ of the respective algebra.
Alternatively they can be obtained from the representation
of the universal R-matrix in case that the underlying symmetry
is a quasi-triangular Hopf algebra.
This is the case for Yangians and quantum affine algebras.
Note that in order to get a rapidity-dependent S-matrix
it is usually not sufficient to demand invariance $\comm{\copro(\gen{J}^A)}{\smat} = 0$
only for $\gen{J}^A\in\alg{g}$, but also for some $\gen{J}^A\in\grp{DY}(\alg{g})$,
since the ordinary Lie generators usually do not
depend on a spectral parameter.
\bigskip

For the AdS/CFT correspondence the symmetry algebra in question is $\alg{psu}(2,2|4)$,
which is broken to $\alg{u}(1)\ltimes\psu\oplus\psu\ltimes\alg{u}(1)$
upon choosing a vacuum state in the Bethe ansatz.
Due to the direct sum structure of this residual symmetry
we can work with one copy of $\psu$.
Interestingly, $\psu$ seems to be the only basic classical Lie superalgebra
which allows for a non-trivial three-dimensional central extension $\psucentral$,
and indeed this central extension seems necessary to derive the S-matrix for our model.
This centrally extended algebra arises both on the gauge \cite{Beisert:2005tm}
and string theory \cite{Arutyunov:2006ak} side of the correspondence.
Interestingly, the S-matrix is already fully fixed \cite{Beisert:2005tm}
up to a factor by demanding only invariance
under the Lie algebra generators of $\psucentral$,
without referring to an additional loop algebra or something similar.
One only needs to introduce an additional braiding element \cite{Gomez:2006va,Plefka:2006ze}
and identify the central charges and the braiding such that the central elements are all cocommutative.
In that case, the S-matrix also depends on spectral parameters.
The reason why it is fully fixed by the Lie algebra generators
lies in the fact that the tensor product
of two fundamental representations of $\psucentral$,
in which the elementary magnons live,
is generically irreducible \cite{Beisert:2006qh}.
Nevertheless, one might wonder if one can lift the $\psucentral$ symmetry
to some loop algebra or one of its deformations.
It has been known since several years that there are some Yangian structures
appearing on both sides of the correspondence \cite{Bena:2003wd,Dolan:2004ps,Dolan:2003uh,Serban:2004jf,Agarwal:2004sz,Zwiebel:2006cb}.
Indeed, in \cite{Beisert:2007ds} it has been shown that the S-matrix
is invariant under the braided Yangian  $\grp{Y}(\psucentral)$.
Recently there have been lots of other activities studying
the encountered algebraic structures, see \cite{Arutyunov:2006yd,Gomez:2007zr,Young:2007wd,Beisert:2007sk}.
Since a Yangian usually has a universal R-matrix, it is natural to ask
if the S-matrix on the fundamental evaluation representation arises
as the representation of this R-matrix.
In particular the overall phase of the S-matrix would directly follow.
Due to quasi-triangularity, which is closely related
to crossing symmetry, see\ \cite{Janik:2006dc},
the dressing factor of the universal R-matrix is constrained and perhaps fully fixed.
This might lead to a derivation of the phase factor proposed in \cite{Beisert:2006ib,Beisert:2006ez}
from first principles.
\bigskip

Even though there are standard methods how to construct the Yangian including its
universal R-matrix for simple Lie algebras \cite{Khoroshkin:1994uk},
there are several reasons why we can not apply them in a simple fashion for our system.
The main one seems to be the peculiar situation that our algebra $\psucentral$
has a non-trivial centre.
This implies that the algebra is not simple and does not even admit
a non-degenerate invariant supersymmetric bilinear form.%
\footnote{One can obtain a non-degenerate form by adjoining 
outer automorphisms, which however do not live on the fundamental representation.}
Then there is the braiding element,
which has to be related to the central charges
in order to have cocommutativity on the centre.
Furthermore, the ordinary Yangian spectral parameter also needs
to be related to the central charges and the braiding.
This makes the situation pretty complicated.
Hence, investigating the classical bialgebra seems a promising way
to study the underlying full quantum Hopf algebra and get an idea
how to obtain its universal R-matrix.
The crucial ingredient for the bialgebra is the classical r-matrix,
which for our system was first investigated in \cite{Torrielli:2007mc}.
In fact, a similar classical r-matrix,
where the momentum scales differently with the coupling constant,
was obtained directly from perturbation theory on the world sheet in \cite{Klose:2006zd},
and in subsequent papers \cite{Klose:2007rz}
the two-loop correction to the classical r-matrix have been computed.
In \cite{Moriyama:2007jt}, an algebraic expression for the classical r-matrix
in the limit performed in \cite{Torrielli:2007mc}
was written down which seems to indicate that the bialgebra
is not the standard loop algebra of $\psu$.
This would probably imply that the universal R-matrix
cannot be obtained by the standard methods.
However, in this paper we argue that the classical r-matrix of $\psucentral$
and the quasi-triangular Lie bialgebra which arises from this classical r-matrix
are almost given by the standard formulae.

\bigskip

We begin by reviewing Lie bialgebras and their relation to Yangian doubles in
\secref{sec:classreview}.
In \secref{sec:su22ext} we apply these methods to $\psucentral$.
We give explicit expressions for the corresponding Lie bialgebra
and its classical r-matrix in \secref{sec:u22def}
and relate it to standard algebras in \secref{sec:relations}.
Different classical limits are investigated in \secref{sec:weak}.
Finally, we provide first steps in lifting the classical bialgebra
structure to the quantum case in \secref{sec:lift}.

%%%%%%%%%%%%%%%%%%%%%%%%%%%%%%%%%%%%%%%%%%%%%%%%%%%%%%%%%%%%%%%%%%%%%%%%%%%%%%%%
%%%%%%%%%%%%%%%%%%%%%%%%%%%%%%%%%%%%%%%%%%%%%%%%%%%%%%%%%%%%%%%%%%%%%%%%%%%%%%%%
\section{Classical r-matrix and Lie Bialgebras}
\label{sec:classreview}

In this section we will review the basic construction of bialgebras,
classical r-matrices and how they arise as limiting cases of certain Hopf algebras.
In particular, we are interested in the bialgebra structure of polynomials
(or Laurent series) with values in a semi-simple Lie algebra,
which lead upon quantisation to (double) Yangians,
which in turn lead to rational solutions of the Yang-Baxter equation (YBE).
We will deal only with the case of simple Lie algebras,
the interested reader will find more details and proofs
for this case in the textbook of Chari and Pressley \cite{Chari:1994pz} 
or in Drinfeld's original report \cite{Drinfeld:1986in}. 
For later purposes let us note that the generalisation
to the case of Lie superalgebras as well as to non-simple Lie algebras
and superalgebras is straightforward provided they allow
for a non-degenerate supersymmetric invariant bilinear form.

%%%%%%%%%%%%%%%%%%%%%%%%%%%%%%%%%%%%%%%%%%%%%%%%%%%%%%%%%%%%%%%%%%%%%%%%%%%%%%%%
\subsection{R-Matrix and Double Yangian}

%%%%%%%%%%%%%%%%%%%%%%%%%%%%%%%%%%%%%%%%
\paragraph{Double Yangian.}

Consider a semi-simple Lie algebra $\alg{g}$
spanned by the generators $\gen{J}^A$
obeying the Lie bracket
\[
\comm{\gen{J}^A}{\gen{J}^B}=F^{AB}_C\gen{J}^C
\]
with the structure constants $F^{AB}_C$.
Define furthermore the Cartan--Killing matrix
$C^{AB}\sim F^{AC}_DF^{BD}_C$,
its inverse $C_{AB}$ and
the conjugated structure constants $F^A_{BC}=F^{AD}_{B}C_{DC}$.

Then the double Yangian $\grp{DY}(\alg{g})$
is a deformation of the universal enveloping
algebra $\grp{U}(\alg{g}[u,u^{-1}])$ of the
loop algebra $\alg{g}[u,u^{-1}]$.
It is generated by the level-$n$ generators $\gen{J}_{n}^A$, $n\in\Integers$,
with level-zero defined to span the Lie algebra, $\gen{J}_{0}^A=\gen{J}^A$.
The commutation relations of these generators read
\[
\comm{\gen{J}^A_m}{\gen{J}^B_n}=F^{AB}_C\gen{J}^C_{n+m}+\order{\hbar},
\]
where $\hbar$ is the deformation parameter.
The precise form of the deformations
for the algebra does not appear very enlightening,
and we have not made it explicit here.
Their coproduct takes the standard form
\<
\copro(\gen{J}_{n}^A)\eq \gen{J}_{n}^A\otimes 1+1\otimes \gen{J}_{n}^A
+\hbar \sum_{m=0}^{n-1}\half  F^A_{BC}\gen{J}_{n-1-m}^B\otimes\gen{J}_{m}^C
+\order{\hbar^2}.
\>
%

%%%%%%%%%%%%%%%%%%%%%%%%%%%%%%%%%%%%%%%%
\paragraph{Universal R-Matrix.}

The double Yangian is quasi-triangular, it has been constructed as a
quantum double of the original Yangian in \cite{Khoroshkin:1994uk}.
This means it has an R-matrix
$\rmat\in\grp{DY}(\alg{g})\otimes\grp{DY}(\alg{g})$
obeying the cocommutativity relation
\[\label{eq:quasicoco}
\coproop(\gen{J}_n^A)\, \rmat
= \rmat\, \copro(\gen{J}_n^A)
\]
with $\coproop:=\perm\circ\copro$ being the opposite coproduct
and $\perm$ the permutation operator. Additional relations ensure that the YBE holds.
We neither spell out these relations nor the explicit form of
the universal R-matrix as we do not need them in what follows.
%In other words the S-matrix $\smat=\perm(\smat)$ is invariant
%under double Yangian
%
%\[
%\comm{\copro(\gen{J}_n^A)}{\perm(\rmat)}=0
%\]
%

%%%%%%%%%%%%%%%%%%%%%%%%%%%%%%%%%%%%%%%%
\paragraph{Evaluation Representations.}

Often one considers evaluation representations of the
Yangian. These representations are most relevant for
integrable spin chains and most transparent.
On a state $\state{u}$ an evaluation representation
of the double Yangian is defined by the action
\[\label{eq:evalrep}
\gen{J}_n^A\state{u}=u^n\gen{J}^A_0\state{u} + \order{\hbar}
\quad\mbox{i.e.}\quad
\gen{J}_n^A\simeq u^n\gen{J}^A_0 + \order{\hbar}.
\]
The representation of $\rmat$
on a state $\state{u_1}\otimes\state{u_2}$ then becomes the
matrix-valued function $R(u_1,u_2)$
which is typically of a difference form $R(u_1-u_2)$,
and leads to rational solutions of YBE.
For invariance of $R(u_1,u_2)$ one merely needs to
check invariance under $\gen{J}^A_0$ and $\gen{J}^A_{1}$
for invariance under $\gen{J}^A_n$ follows from an identity
\[\label{eq:doubleyang}
\copro(\gen{J}_{n}^A)
\simeq
\frac{u_1^{n-1}-u_2^{n-1}}{u_1^{-1}-u_2^{-1}}\,\copro(\gen{J}_0^A)
+\frac{u_1^{n}-u_2^{n}}{u_1-u_2}\,\copro(\gen{J}_{1}^A) + \order{\hbar}
\]
which holds for evaluation representations.

%%%%%%%%%%%%%%%%%%%%%%%%%%%%%%%%%%%%%%%%%%%%%%%%%%%%%%%%%%%%%%%%%%%%%%%%%%%%%%%%
\subsection{Classical Limit and Lie Bialgebra}

%%%%%%%%%%%%%%%%%%%%%%%%%%%%%%%%%%%%%%%%
\paragraph{Classical Limit.}

Now let us consider the classical limit of the above
algebra where we restrict to the first order in $\hbar$ everywhere.
We first expand the coproduct and opposite coproduct
\[
\copro=\copro_0+\hbar\copro_1+\order{\hbar^2},\qquad
\coproop=\copro_0+\hbar\coproop_1+\order{\hbar^2}.
\]
The classical r-matrix is obtained from the
quantum R-matrix by expansion in the
deformation parameter $\hbar$
\[
\rmat=1\otimes 1+\hbar\,r+\order{\hbar}^2.
\]
By substituting these two expressions into the
quasi-cocommutativity \eqref{eq:quasicoco}
relation we obtain
\[\label{eq:limitquasicoco}
\comm{\copro_0(\gen{J}_n^A)}{r}
=
\copro_1(\gen{J}_n^A)-\coproop_1(\gen{J}_n^A).
\]
Similarly, if $\rmat$ satisfies the
quantum YBE $\rmat_{12}\rmat_{13}\rmat_{23} = \rmat_{23}\rmat_{13}\rmat_{12} $,
it is straightforward to check that the classical r-matrix will satisfy
\[\label{eq:cybe}
\cybe{r}{r} := \comm{r_{12}}{r_{13}} + \comm{r_{12}}{r_{23}} + \comm{r_{13}}{r_{23}} = 0,
\]
which is called the classical Yang-Baxter equation (CYBE).

%%%%%%%%%%%%%%%%%%%%%%%%%%%%%%%%%%%%%%%%
\paragraph{Lie Bialgebras.}

The above expansion can be cast into the framework of a Lie bialgebra.
In general, a Lie bialgebra is a Lie algebra $\alg{g}$ equipped
with an antisymmetric linear map,
called the \emph{cobracket},
\[
\coprocl :\alg{g} \to \alg{g}\otimes\alg{g},
\]
such that the dual map $\coprocl^* :\alg{g}^*\otimes\alg{g}^* \to \alg{g}^*$
is an ordinary Lie bracket. This means that if $(F^*)_{AB}{}^C$ are structure constants of the cobracket, i.e. $\coprocl{\gen{J}^A} = (F^*)_{BC}{}^A \gen{J}^B\otimes\gen{J}^C$, then the same constants define the Lie bracket of $\alg{g}^*$ for the corresponding dual basis, $\comm{\gen{J}_A}{\gen{J}_B} = (F^*)_{AB}{}^C\gen{J}_C$. Similarly, the structure constants of $\alg{g}$ define a cobracket on $\alg{g}^*$.
Furthermore the cobracket $\coprocl$ is a cocycle, i.e.
\[
\coprocl(\comm{\gen{J}_1}{\gen{J}_2})
= \comm{\gen{J}_1}{\coprocl(\gen{J}_2)} - \comm{\gen{J}_2}{\coprocl(\gen{J}_1)} ,
\]
where one extends the
Lie bracket canonically to the space
$\alg{g}\otimes \alg{g}$
by defining (with proper signs due to fermi statistics implicit)
\[\label{eq:ggBracket}
\comm{\gen{J}_1}{\gen{J}_2\otimes\gen{J}_3}
=-\comm{\gen{J}_2\otimes\gen{J}_3}{\gen{J}_1}
:=
\comm{\gen{J}_1}{\gen{J}_2}\otimes\gen{J}_3
+\gen{J}_2\otimes\comm{\gen{J}_1}{\gen{J}_3}.
\]

We will be especially interested in \emph{coboundary} bialgebras
where the cobracket is obtained by commuting with a classical r-matrix $\crmat$,
\[\label{eq:classcoco}
\coprocl(\gen{J}) = \comm{\gen{J}}{\crmat}.
\]
The properties of the Lie bialgebra are
satisfied if $\cybe{r}{r}$, cf.\ \eqref{eq:cybe},
is invariant under the Lie algebra.
In particular, if the r-matrix satisfies
the classical Yang-Baxter equation
$\cybe{r}{r}=0$
the Lie bialgebra is called \emph{quasi-triangular}.

The relation \eqref{eq:classcoco} matches equation \eqref{eq:limitquasicoco}
if we relate the cobracket as
\[\label{eq:coprocobra}
\coprocl(\gen{J}_n^A)=\copro_1(\gen{J}_n^A)-\coproop_1(\gen{J}_n^A)
\]
for according to
\eqref{eq:ggBracket}
we have
$\comm{\copro_0(\gen{J}_n^A)}{r}=\comm{\gen{J}_n^A}{r}$.
Note that our cobracket \eqref{eq:coprocobra} is obviously antisymmetric.
Thus we have now formulated the relations in the classical limit
purely in terms of a quasi-triangular Lie bialgebra.

In fact, what we presented in this paragraph is quite generic: whenever we
have a quasi-triangular Hopf algebra which is a
deformation of a universal enveloping algebra, 
we obtain a corresponding
quasi-triangular Lie bialgebra by considering 
the properties of the Hopf algebra 
at the lowest orders in the deformation parameter $\hbar$, 
see \cite{Chari:1994pz}.

%%%%%%%%%%%%%%%%%%%%%%%%%%%%%%%%%%%%%%%%
\paragraph{Loop Algebra.}

The classical limit of a double Yangian leads to a Lie bialgebra based on the
loop algebra $\alg{g}[u,u^{-1}]$ of $\alg{g}$.
It has the standard bracket
\[
\comm{\gen{J}^A_m}{\gen{J}^B_n}=F^{AB}_C\gen{J}^C_{n+m}.
\]
The cobracket is defined as
\[
\coprocl(\gen{J}_n^A)
=\half F^A_{BC}\sum_{m=0}^{n-1}\gen{J}_{n-1-m}^B\wedge\gen{J}_{m}^C,
\]
with the antisymmetric tensor product
\[
\gen{J}_1\wedge \gen{J}_2:=\gen{J}_1\otimes \gen{J}_2-\gen{J}_2\otimes \gen{J}_1.
\]
It is not hard to confirm that the antisymmetric classical r-matrix
\[\label{eq:standardrantisym}
r=\sum_{n=0}^\infty\half C_{CD}
\gen{J}_{-1-n}^C\wedge\gen{J}_{n}^D
\]
obeys the relation \eqref{eq:classcoco}.
Furthermore, $\cybe{r}{r}$ is algebra-invariant
and therefore the bialgebra is coboundary.
In order to satisfy the CYBE $\cybe{r}{r}=0$
we have to choose an asymmetric form for the r-matrix
\[\label{eq:standardrasym}
r=
\sum_{n=0}^\infty C_{CD}
\gen{J}_{-1-n}^C\otimes\gen{J}_{n}^D
\qquad\mbox{or}\qquad
r=
-\sum_{n=0}^\infty C_{CD}
\gen{J}_{n}^C\otimes\gen{J}_{-1-n}^D.
\]
In both cases the Lie bialgebra is quasi-triangular.

%%%%%%%%%%%%%%%%%%%%%%%%%%%%%%%%%%%%%%%%
\paragraph{Evaluation Representations.}

Consider now evaluation representations as above in
\eqref{eq:evalrep}.
The representation of the r-matrix
on a state $\state{u_1}\otimes\state{u_2}$ becomes
\[
r\simeq
r(u_1,u_2)=\frac{C_{CD}\gen{J}_0^C\otimes\gen{J}_0^D}{u_1-u_2}\,.
\]
In fact all three forms \eqref{eq:standardrantisym,eq:standardrasym}
are equivalent up to contact terms at $u_1=u_2$.
The above action is proportional to the quadratic Casimir operator
of the Lie algebra at level-0 and therefore obviously
\[
\comm{\gen{J}_0^A}{r}\simeq0
=\coprocl(\gen{J}_0^A).
\]
For the level-one generator $\gen{J}_{1}^A$ one also finds that
the coboundary relation holds
\[\comm{\gen{J}_1^A}{r}
\simeq
F^{A}_{BC}\gen{J}_0^B\otimes\gen{J}_0^C
=\coprocl(\gen{J}_{1}^A).
\]
For evaluation representations,
the coboundary property for the remaining generator $\gen{J}_n^A$
follows from the level-zero and level-one relations
by means of the identity
\[\label{eq:cldoubleyang}
\coprocl(\gen{J}_{n}^A)
\simeq
\frac{u_1^{n-1}-u_2^{n-1}}{u_1^{-1}-u_2^{-1}}\,\coprocl(\gen{J}_0^A)
+\frac{u_1^{n}-u_2^{n}}{u_1-u_2}\,\coprocl(\gen{J}_{1}^A).
\]

%%%%%%%%%%%%%%%%%%%%%%%%%%%%%%%%%%%%%%%%%%%%%%%%%%%%%%%%%%%%%%%%%%%%%%%%%%%%%%%%
%%%%%%%%%%%%%%%%%%%%%%%%%%%%%%%%%%%%%%%%%%%%%%%%%%%%%%%%%%%%%%%%%%%%%%%%%%%%%%%%
\section{Centrally Extended $\alg{su}(2|2)$}
\label{sec:su22ext}

%%%%%%%%%%%%%%%%%%%%%%%%%%%%%%%%%%%%%%%%%%%%%%%%%%%%%%%%%%%%%%%%%%%%%%%%%%%%%%%%
\subsection{Yangian Double}

We would like to understand the R-matrix
that appears in the context of AdS/CFT
on an algebraic level.
Its symmetry is based on centrally extended
$\alg{su}(2|2)$ symmetry \cite{Beisert:2004ry,Beisert:2005tm}
\[\alg{h}:=\alg{su}(2|2)\ltimes\Reals^2=\alg{psu}(2|2)\ltimes\Reals^3\]
and it acts on two four-dimensional fundamental representations
of $\alg{h}$.
It is generated by the
$\alg{su}(2)\times\alg{su}(2)$ generators $\gen{R}^{a}{}_{b}$,
$\gen{L}^{\alpha}{}_{\beta}$, the supercharges
$\gen{Q}^{\alpha}{}_{b}$, $\gen{S}^{a}{}_{\beta}$ and the central
charges $\gen{C}$, $\gen{P}$, $\gen{K}$.
Where appropriate, we shall use the
collective symbol $\gen{J}^A$ for these generators.
The R-matrix also displays Yangian symmetry $\grp{Y}(\alg{h})$ \cite{Beisert:2007ds}
and by means of \eqref{eq:doubleyang}
double Yangian symmetry $\grp{DY}(\alg{h})$.
The level-$n$ generators corresponding
to $\gen{J}^A$ shall be denoted by $\gen{J}_n^A$.

%%%%%%%%%%%%%%%%%%%%%%%%%%%%%%%%%%%%%%%%
\paragraph{Commutators.}

The Lie brackets of the $\alg{su}(2)\times\alg{su}(2)$ generators
take the standard form
\[\label{eq:su2su2}
\begin{array}[b]{rclcrcl}
\comm{\gen{R}^a{}_b}{\gen{R}^c{}_d}\eq
\delta^c_b\gen{R}^a{}_d-\delta^a_d\gen{R}^c{}_b,
&&
\comm{\gen{L}^\alpha{}_\beta}{\gen{L}^\gamma{}_\delta}\eq
\delta^\gamma_\beta\gen{L}^\alpha{}_\delta-\delta^\alpha_\delta\gen{L}^\gamma{}_\beta,
\\[3pt]
\comm{\gen{R}^a{}_b}{\gen{Q}^\gamma{}_d}\eq
-\delta^a_d\gen{Q}^\gamma{}_b+\half \delta^a_b\gen{Q}^\gamma{}_d,
&&
\comm{\gen{L}^\alpha{}_\beta}{\gen{Q}^\gamma{}_d}\eq
+\delta^\gamma_\beta\gen{Q}^\alpha{}_d-\half \delta^\alpha_\beta\gen{Q}^\gamma{}_d,
\\[3pt]
\comm{\gen{R}^a{}_b}{\gen{S}^c{}_\delta}\eq
+\delta^c_b\gen{S}^a{}_\delta-\half \delta^a_b\gen{S}^c{}_\delta,
&&
\comm{\gen{L}^\alpha{}_\beta}{\gen{S}^c{}_\delta}\eq
-\delta^\alpha_\delta\gen{S}^c{}_\beta+\half \delta^\alpha_\beta\gen{S}^c{}_\delta.
\end{array}\]
The Lie brackets of two supercharges yield
\<\label{eq:QQ}
\acomm{\gen{Q}^\alpha{}_b}{\gen{S}^c{}_\delta}\eq
\delta^c_b\gen{L}^\alpha{}_\delta +\delta^\alpha_\delta\gen{R}^c{}_b
+\delta^c_b\delta^\alpha_\delta\gen{C},
\nln
\acomm{\gen{Q}^{\alpha}{}_{b}}{\gen{Q}^{\gamma}{}_{d}}\eq
\varepsilon^{\alpha\gamma}\varepsilon_{bd}\gen{P},
\nln
\acomm{\gen{S}^{a}{}_{\beta}}{\gen{S}^{c}{}_{\delta}}\eq
\varepsilon^{ac}\varepsilon_{\beta\delta}\gen{K}.
\>
The remaining Lie brackets vanish.
Again, we do not write the commutators
of the level-one generators explicitly.

%%%%%%%%%%%%%%%%%%%%%%%%%%%%%%%%%%%%%%%%
\paragraph{Coproduct.}

For the coproduct one should introduce a non-trivial braiding
\cite{Gomez:2006va,Plefka:2006ze,Beisert:2007ds}
\<\label{eq:braid}
\copro(\gen{J}_n^A)\eq
\gen{J}_n^A\otimes 1+\eip^{[A]}\otimes\gen{J}_n^A
+\half g^{-1} F^{A}_{BC}\sum_{k=0}^{n-1}\gen{J}_k^B\eip^{[C]}\otimes\gen{J}_{n-1-k}^C
+\order{g^{-2}},
\nln
\copro(\eip)\eq \eip\otimes\eip.
\>
with some abelian generator $\eip$ (a priori unrelated to the algebra)
and the grading
\[\label{eq:grading}
[\gen{K}]=-2,\quad
[\gen{S}]=-1,\quad
[\gen{R}]=[\gen{L}]=[\gen{C}]=0,\quad
[\gen{Q}]=+1,\quad
[\gen{P}]=+2.
\]
This ``braid charge'' is proportional to the
charge under the external $\alg{u}(1)$ automorphism $\bar{\gen{B}}$ acting as
$\comm{\bar{\gen{B}}}{\gen{J}^A}=[A]\gen{J}^A$,
and therefore the coproduct is compatible with the algebra relations.

To achieve a quasi-cocommutative algebra,
the central charges $\gen{P}_0,\gen{K}_0,\gen{P}_1,\gen{K}_1$ must be identified
with the braiding factor $\eip$
and the central charge $\gen{C}_0$ as follows \cite{Plefka:2006ze,Beisert:2007ds}%
\footnote{We set the inessential shift parameter $u_0$ in
\protect\cite{Beisert:2007ds} to zero.}
\[\label{eq:PKident}
\begin{array}[b]{rclcrcl}
\gen{P}_0\eq g\alpha\bigbrk{1-\eip^{+2}},
&&
\gen{K}_0\eq g\alpha^{-1}\bigbrk{1-\eip^{-2}},
\\[3pt]
\gen{P}_1\eq\alpha\gen{C}_0\bigbrk{1+\eip^{+2}},
&&
\gen{K}_1\eq-\alpha^{-1}\gen{C}_0\bigbrk{1+\eip^{-2}}.
\end{array}
\]
{}From these identifications it follows that
the evaluation parameter $iu$ for
any evaluation representation
is fixed in terms of the eigenvalues of $\gen{C}$ and $\eip$
\[
\gen{J}_n^A \simeq
(iu)^n \gen{J}_0^A ,\qquad
iu\simeq g^{-1} \gen{C}_0\,\frac{1+\eip^{+2}}{1-\eip^{+2}}\,.
\]
%

%%%%%%%%%%%%%%%%%%%%%%%%%%%%%%%%%%%%%%%%
\paragraph{Fundamental Representation.}

The algebra $\alg{h}$ has a four-dimensional
representation \cite{Beisert:2005tm}
which we will call fundamental.
The corresponding multiplet has two bosonic states $\state{\phi^a}$
and two fer\-mi\-onic states $\state{\psi^\alpha}$.
The action of the two sets of $\alg{su}(2)$ generators
has to be canonical
\[\label{eq:bosegen}
\gen{R}^a{}_b\state{\phi^c}=\delta^c_b\state{\phi^a}
  -\half \delta^a_b\state{\phi^c},
\qquad
\gen{L}^\alpha{}_\beta\state{\psi^\gamma}=\delta^\gamma_\beta\state{\psi^\alpha}
  -\half \delta^\alpha_\beta\state{\psi^\gamma}.
\]
The supersymmetry generators must also act in a manifestly
$\alg{su}(2)\times\alg{su}(2)$ covariant way
\[\label{eq:fermigen}
\begin{array}[b]{rclcrcl}
\gen{Q}^\alpha{}_a\state{\phi^b}\eq a\,\delta^b_a\state{\psi^\alpha},&&
\gen{Q}^\alpha{}_a\state{\psi^\beta}\eq b\,\varepsilon^{\alpha\beta}\varepsilon_{ab}\state{\phi^b},\\[3pt]
\gen{S}^a{}_\alpha\state{\phi^b}\eq c\,\varepsilon^{ab}\varepsilon_{\alpha\beta}\state{\psi^\beta},&&
\gen{S}^a{}_\alpha\state{\psi^\beta}\eq d\,\delta^\beta_\alpha\state{\phi^a}.
\end{array}\]
We can write the four parameters $a,b,c,d$
using the parameters $\xpm{}$, $\gamma$ and the constants $g$, $\alpha$ as
\[\label{eq:abcdQ}
a=\sqrt{g}\,\gamma,\quad
b=\sqrt{g}\,\frac{\alpha}{\gamma}\lrbrk{1-\frac{\xp{}}{\xm{}}},\quad
c=\sqrt{g}\,\frac{i\gamma}{\alpha \xp{}}\,,\quad
d=\sqrt{g}\,\frac{\xp{}}{i\gamma}\lrbrk{1-\frac{\xm{}}{\xp{}}}.
\]
The parameters $\xpm{}$ (together with $\gamma$) label
the representation%
\footnote{For a hermitian representation we should set
$|\gamma|=|\sqrt{-i\xp{}+i\xm{}}|$ and $|\alpha|=1$.}
and they must obey the constraint
\[\label{eq:xpmIdent}
\xp{}+\frac{1}{\xp{}}-\xm{}-\frac{1}{\xm{}}=\frac{i}{g}\,.
\]
The three central charges $\gen{C},\gen{P},\gen{K}$ and $\eip$
are represented by the values $C,P,K$ and $U$ which read
\[
C=\frac{1}{2}\,\frac{1+1/\xp{}\xm{}}{1-1/\xp{}\xm{}}\,,
\quad
P=g\alpha\lrbrk{1-\frac{\xp{}}{\xm{}}},
\quad
K=\frac{g}{\alpha}\lrbrk{1-\frac{\xm{}}{\xp{}}},
\quad
U=e^{ip/2}=\sqrt{\frac{\xp{}}{\xm{}}}\,.
\]
The coefficient $U$ is most immediately related
to the particle momentum $p$ used in the scattering matrix by $U=e^{ip/2}$.
These eigenvalues obey the quadratic relation $C^2-PK=\quarter$
by virtue of \eqref{eq:xpmIdent}.
Note that the corresponding quadratic combination of central charges
$\gen{C}^2-\gen{P}\gen{K}$ is singled out by being
invariant under the external $\alg{sl}(2)$ automorphism
of $\alg{h}$, see \secref{sec:sl2auto}.

The representation of Yangian $\grp{DY}(\alg{h})$ is of evaluation type
$\gen{J}^A_n\simeq (iu)^n \gen{J}^A_0$ \cite{Beisert:2007ds}.
The evaluation parameter $u$ is related to the $\xpm{}$ parameters
by
\[\label{eq:relxpmu}
u=\xp{}+\frac{1}{\xp{}}-\frac{i}{2g}=\xm{}+\frac{1}{\xm{}}+\frac{i}{2g}
=\half(\xp{}+\xm{})(1+1/\xp{}\xm{})\,.
\]
%

%%%%%%%%%%%%%%%%%%%%%%%%%%%%%%%%%%%%%%%%
\paragraph{Fundamental R-Matrix.}

In \cite{Beisert:2005tm,Beisert:2006qh}
an S-matrix acting on the tensor product of
two fundamental representations was derived.
It was constructed by imposing invariance under
the algebra $\alg{h}$ \cite{Beisert:2005tm,Beisert:2006qh},
and it was shown to be invariant under Yangian
generators \cite{Beisert:2007ds}
\[\label{eq:Sinv}
\comm{\copro(\gen{J}^A_n)}{\smat}=0.
\]
The S-matrix also satisfies the YBE \cite{Beisert:2005tm,Beisert:2006qh}.
We will not reproduce the result here, it is given in \cite{Beisert:2006qh}.
Note that we have to fix the parameters $\xi=U=\sqrt{\xp{}/\xm{}}$
in order to make the action of the generators
compatible with the coproduct \eqref{eq:braid}.%
\footnote{This identification removes all
braiding factors from the S-matrix
in \protect\cite{Beisert:2006qh} which will
thus satisfy the standard Yang-Baxter (matrix) equation,
see also \protect\cite{Beisert:2005tm,Arutyunov:2006yd,Martins:2007hb}.}

From the S-matrix we can read off a fundamental R-matrix
\[\smat=\perm \rmat,\]
where $\perm$ is a (graded) permutation operator.
Upon this identification,
invariance of the S-matrix in \eqref{eq:Sinv}
is equivalent to quasi-cocommutativity \eqref{eq:quasicoco} of the R-matrix.

The next step would be to construct the universal R-matrix
for the algebra $\alg{h}$.
Our centrally extended algebra $\alg{h}$
is however not semi-simple and therefore the standard
construction of the universal R-matrix cannot be applied.
The main reason for the failure is
that the Cartan--Killing matrix $C^{AB}$ is singular
and its inverse $C_{AB}$,
which plays an important role in the construction, does not exist.
Similarly, for the standard construction of the classical r-matrix
one would need the quadratic Casimir $\half C_{AB}\gen{J}^A\gen{J}^B$
which does not exist for our algebra.%
\footnote{Due to the deformed coproduct for Lie generators,
this is actually not what one really wants.}

The R-matrix has one overall phase factor $S^0$
\[\label{eq:phaseexact}
S^0_{12} = \exp(i\theta_{21})\,
\sqrt{
\frac{1-1/\xp{2}\xm{1}}{1-1/\xm{2}\xp{1}}\,\frac{\xm{2}-\xp{1}}{\xp{2}-\xm{1}}}\,,
\]
where $\theta$ is the so-called dressing phase.
The phase cannot be determined from quasi-cocommutativity.
Quasi-triangularity, however, imposes some constraint
which is believed to give the crossing symmetry relation found in \cite{Janik:2006dc}.
In \cite{Beisert:2006ib,Beisert:2006ez} a proposal for a crossing-symmetric phase was made.
The proposal is fully consistent with perturbative results from gauge theory \cite{Serban:2004jf}
and from string theory \cite{Arutyunov:2004vx,Beisert:2005cw,Hernandez:2006tk,Gromov:2007aq}.

For simplicity we shall choose a specific dressing factor
which does not obey crossing.
It turns out that the light cone string S-matrix \cite{Frolov:2006cc}
leads to convenient and symmetric expressions.
The dressing factor for this case is
\[\label{eq:phaseLC}
S^0_{12} = \sqrt{\sqrt{\frac{\xp{1}\xm{2}}{\xm{1}\xp{2}}}\,
\frac{\xm{2}-\xp{1}}{\xp{2}-\xm{1}}}\,,
\qquad
(A_{12})^2=\sqrt{\frac{\xp{1}\xm{2}}{\xm{1}\xp{2}}}\,\frac{\xp{2}-\xm{1}}{\xm{2}-\xp{1}}\,.
\]
At leading order at strong coupling
(on which we will focus our attention in the remainder of the paper)
this phase factor agrees with the correct physical result up
to a term which can be absorbed into the definition of the length of the string.
Another useful choice is
\[\label{eq:phaseexactmom}
S^0_{12} = \exp(i\theta_{21})\,
\sqrt{
\frac{\xp{1}\xm{2}}{\xm{1}\xp{2}}
\,\frac{1-1/\xp{2}\xm{1}}{1-1/\xm{2}\xp{1}}\,\frac{\xm{2}-\xp{1}}{\xp{2}-\xm{1}}}\,,
\]
which differs from \eqref{eq:phaseexact}
by some factors of the particle momenta
leading to a redefinition of the length of a state.

%%%%%%%%%%%%%%%%%%%%%%%%%%%%%%%%%%%%%%%%%%%%%%%%%%%%%%%%%%%%%%%%%%%%%%%%%%%%%%%%
\subsection{Classical Limit and Lie Bialgebra}

A suitable classical limit for the above S-matrix
the limit where the particle momenta $p$ scale like $\hbar$
while the coupling constant $g$ approaches infinity like $1/\hbar$.
This limit is well-known as the (near) plane wave limit
\cite{Berenstein:2002jq,Parnachev:2002kk,Callan:2003xr,Minahan:2002ve}
(for finitely many excitations above the vacuum)
or as the classical limit for spinning strings
\cite{Frolov:2003qc,Beisert:2003xu,Kruczenski:2003gt,Kazakov:2004qf}
(for coherent states of infinitely many excitations).
In this limit the evaluation parameter $u$ becomes
large as $1/\hbar$ as for typical classical limits.

One may also consider a different limit $u\sim 1/\hbar$
but $g\sim 1/\hbar^\kappa$ with adjustable $\kappa$.
The standard classical limit corresponds to $\kappa=1$.
For $\kappa>1$ it turns out that limit of the R-matrix
is not of the form $\rmat=1\otimes 1+\order{\hbar}$.
Conversely, for $\kappa<1$ the R-matrix has a classical limit
$\rmat=1\otimes 1+\order{\hbar}$, 
but with an r-matrix which is a twisted version of the
standard $\alg{u}(2|2)$ r-matrix.
We shall review this case in \secref{sec:weak}.

%%%%%%%%%%%%%%%%%%%%%%%%%%%%%%%%%%%%%%%%
\paragraph{Lie Bialgebra.}

The classical limit described above is the limit $g\to\infty$,
i.e.\ the quantum parameter is $\hbar=g^{-1}$,
while assuming that
\[
\eip=\exp (\ihalf g^{-1} \gen{D})
\]
with some finite abelian generator $\gen{D}$.
This ensures that the charges $\gen{P},\gen{K}$ remain finite
in the limit:
\[
\gen{P}=-i\alpha\gen{D}+\order{g^{-1}},\qquad
\gen{K}=i\alpha^{-1}\gen{D}+\order{g^{-1}}.
\]
The fundamental R-matrix becomes trivial in the limit and
the first perturbation yields the classical r-matrix
\[
\rmat=1\otimes 1+g^{-1}r+\order{g^{-2}}.
\]

In the classical limit, the Lie brackets of two supercharges
\eqref{eq:QQ} read
\<
\acomm{\gen{Q}^\alpha{}_b}{\gen{S}^c{}_\delta}\eq
\delta^c_b\gen{L}^\alpha{}_\delta +\delta^\alpha_\delta\gen{R}^c{}_b
+\delta^c_b\delta^\alpha_\delta\gen{C},
\nln
\acomm{\gen{Q}^{\alpha}{}_{b}}{\gen{Q}^{\gamma}{}_{d}}\eq
-i\alpha \varepsilon^{\alpha\gamma}\varepsilon_{bd}\gen{D},
\nln
\acomm{\gen{S}^{a}{}_{\beta}}{\gen{S}^{c}{}_{\delta}}\eq
i\alpha^{-1}\varepsilon^{ac}\varepsilon_{\beta\delta}\gen{D},
\>
i.e.\ the abelian generator $\gen{D}$ replaces
$\gen{P}$ and $\gen{K}$ and becomes part of the
Lie algebra.
The limit of the coproduct \eqref{eq:braid} yields the cobrackets
via \eqref{eq:coprocobra}
\[\label{eq:cobraextsu22}
\coprocl(\gen{J}_n^A)=
\ihalf [A]\,\gen{D}\wedge\gen{J}_n^A
+\half F^{A}_{BC}\sum_{k=0}^{n-1}\gen{J}_k^B\wedge\gen{J}_{n-1-k}^C.
\]
The grading for the new generator $\gen{D}$ is obviously
trivial $[\gen{D}]=0$.

{}From these identifications it follows that
the evaluation parameter $iu$ for
any evaluation representation
is fixed in terms of the eigenvalues of $\gen{C}$ and $\gen{D}$
\[\label{eq:spectral}
\gen{J}_n^A \simeq
(iu)^n \gen{J}^A ,\qquad
u\simeq 2\gen{C}\gen{D}^{-1}\,.
\]

%%%%%%%%%%%%%%%%%%%%%%%%%%%%%%%%%%%%%%%%
\paragraph{Fundamental Representation.}

The fundamental representation simplifies somewhat in the
classical limit. We choose the following parametrisation \cite{Arutyunov:2006iu}
for the kinematical variables $\xpm{}$ :
\[
\xpm{}=x\sqrt{1-\frac{1}{4g^2(x-1/x)^2}}\pm \frac{i}{2g}\,\frac{x}{x-1/x}\,,\qquad
\gamma = \frac{1}{\sqrt{g}}\,\gat{}
\]
The parameters $\gat{},x,\alpha$ are independent of $g$.

The action of the Lie generators on the $2|2$-dimensional representation
space spanned by $\state{\phi^a}$ and $\state{\psi^\alpha}$
was given in \eqref{eq:bosegen,eq:fermigen}.
The limit of the coefficients $a,b,c,d$ are given as follows:
\[\label{eq:abcd}
a = \gat{},\qquad
b = -\frac{i\alpha x}{\gat{}(x^2-1)}\,,\qquad
c = \frac{i\gat{}}{\alpha x}\,,\qquad
d = \frac{x^2}{\gat{}(x^2-1)}\,.
\]
The eigenvalues of the central charges $\gen{D}$ and $\gen{C}$
read
\[
D = \frac{x}{x^2-1}\,, \qquad
C = \frac{1}{2}\,\frac{x^2+1}{x^2-1}\,.
\]
Furthermore, the Yangian spectral parameter $u$ is simply given
by $u = x + 1/x$, so we immediately confirm \eqref{eq:spectral},
\[
u = x + \frac{1}{x} = 2CD^{-1}.
\]
Finally let us mention that for a hermitian representation
we should put $\alpha=1$, $\gat{}=\sqrt{xD}$.

%%%%%%%%%%%%%%%%%%%%%%%%%%%%%%%%%%%%%%%%
\paragraph{Fundamental r-matrix.}

Let us now take the classical limit of the R-matrix in
the fundamental representation using $\rmat = 1\otimes 1 + g^{-1}r$.
The resulting representation of the classical r-matrix
is given in \tabref{tab:rCoeff}, see \cite{Torrielli:2007mc}.%
\footnote{One should be able to read off the elements of the diagonalised r-matrix
from the integral kernels in  \cite{Beisert:2005bm}.}
The choice of the phase corresponds to the light cone string S-matrix
\cite{Frolov:2006cc} in \eqref{eq:phaseLC}.
Instead one could also choose the exact
phase factor \eqref{eq:phaseexactmom}
obtained in \cite{Arutyunov:2004vx} at the classical level
(with a redefinition of length).
In that case one would have to add the
term $(1\times 1)r_0$ to $r$ with
\[\label{eq:AFSphaseshift}
r_0\simeq
\frac{i(x_1-x_2)(x_1x_2-1)}{4(x_1^2-1)(x_2^2-1)}
=\quarter(iu_2-iu_1)D_1D_2
=(1/iu_2-1/iu_1)C_1C_2.
\]
\begin{table}
\<
\crmat\state{\phi^a\phi^b}\eq
\half (A_{12}-B_{12})\state{\phi^a\phi^b}
+\half (A_{12}+B_{12})\state{\phi^b\phi^a}
+\half C_{12}\varepsilon^{ab}\varepsilon_{\alpha\beta}\state{\psi^\alpha\psi^\beta}
\nln
\crmat\state{\psi^\alpha\psi^\beta}\eq
-\half (D_{12}-E_{12})\state{\psi^\alpha\psi^\beta}
-\half (D_{12}+E_{12})\state{\psi^\beta\psi^\alpha}
-\half F_{12}\varepsilon^{\alpha\beta}\varepsilon_{ab}\state{\phi^a\phi^b}
\nln
\crmat\state{\phi^a\psi^\beta}\eq
G_{12}\state{\phi^a\psi^\beta}
+H_{12}\state{\psi^\beta\phi^a}
\nln
\crmat\state{\psi^\alpha\phi^b}\eq
K_{12}\state{\phi^b\psi^\alpha}
+L_{12}\state{\psi^\alpha\phi^b}
\nonumber
\>

\<
\half(A_{12}+B_{12})\eq\frac{1}{iu_1-iu_2}\nln
\half(A_{12}-B_{12})\eq\frac{(x_1-x_2)^2(x_1x_2+1)^2}{4x_1x_2(x_1^2-1)(x^2_2-1)(iu_1-iu_2)}
      =\frac{+\half+\quarter D_1D^{-1}_2+\quarter D^{-1}_1D_2}{iu_1-iu_2}\nln
\half C_{12}\eq \frac{i\gat{1}\gat{2}(x_1-x_2)}{\alpha x_1x_2(iu_1-iu_2)}
      =\frac{a_1c_2-c_1a_2}{iu_1-iu_2}\nln
-\half(D_{12}+E_{12})\eq-\frac{1}{iu_1-iu_2}\nln
-\half(D_{12}-E_{12})\eq-\frac{(x_1-x_2)^2(x_1x_2+1)^2}{4x_1x_2(x_1^2-1)(x^2_2-1)(iu_1-iu_2)}
      =\frac{-\half-\quarter D_1D^{-1}_2-\quarter D^{-1}_1D_2}{iu_1-iu_2}\nln
-\half F_{12}\eq -\frac{i\alpha x_1x_2(x_1-x_2)}{\gat{1}\gat{2}(x_1^2-1)(x^2_2-1)(iu_1-iu_2)}
      =\frac{d_1b_2-b_1d_2}{iu_1-iu_2}\nln
G_{12}\eq \frac{(x^2_1-x^2_2)(x^2_1x^2_2-1)}{4x_1x_2(x_1^2-1)(x_2^2-1)(iu_1-iu_2)}
      =\frac{-\quarter D_1D^{-1}_2+\quarter D^{-1}_1D_2}{iu_1-iu_2}\nln
H_{12}\eq \frac{\gat{1} x_2(x_1x_2-1)}{\gat{2} x_1(x_2^2-1)(iu_1-iu_2)}
      =\frac{a_1d_2-c_1b_2}{iu_1-iu_2}\nln
K_{12}\eq \frac{\gat{2}x_1(x_1x_2-1)}{\gat{1}x_2(x_1^2-1)(iu_1-iu_2)}
      =\frac{d_1a_2-b_1c_2}{iu_1-iu_2}\nln
L_{12}\eq -\frac{(x^2_1-x^2_2)(x^2_1x^2_2-1)}{4x_1x_2(x_1^2-1)(x_2^2-1)(iu_1-iu_2)}
      =\frac{+\quarter D_1D^{-1}_2-\quarter D^{-1}_1D_2}{iu_1-iu_2}\nonumber
\>
\caption{The classical (light cone) r-matrix of AdS/CFT.}
\label{tab:rCoeff}
\end{table}

%%%%%%%%%%%%%%%%%%%%%%%%%%%%%%%%%%%%%%%%%%%%%%%%%%%%%%%%%%%%%%%%%%%%%%%%%%%%%%%%
%%%%%%%%%%%%%%%%%%%%%%%%%%%%%%%%%%%%%%%%%%%%%%%%%%%%%%%%%%%%%%%%%%%%%%%%%%%%%%%%
\section{A Lie Bialgebra for the Classical r-Matrix}
\label{sec:u22def}

%%%%%%%%%%%%%%%%%%%%%%%%%%%%%%%%%%%%%%%%%%%%%%%%%%%%%%%%%%%%%%%%%%%%%%%%%%%%%%%%
\subsection{Moriyama--Torrielli proposal for the classical r-matrix}
\label{sec:MT}

In \cite{Moriyama:2007jt} the following expression for the classical r-matrix has been proposed:
\<
 r\indup{MT} \eq \sum_{m=0}^{\infty}\Bigl[
+(\gen{R}_m)^a{}_{b}\otimes(\gen{\tilde R }_{-1-m})^b{}_{a}
- (\gen{L}_m)^\alpha{}_{\beta}\otimes(\gen{\tilde L}_{-1-m})^\beta{}_{\alpha}
\nl\qquad
- (\gen{R}_{-1-m})^a{}_{b}\otimes(\gen{\tilde R}_m)^b{}_{a}
+ (\gen{L}_{-1-m})^\alpha{}_{\beta}\otimes(\gen{\tilde L}_{m})^\beta{}_{\alpha}
\nl\qquad
+(\gen{Q}_m)^\alpha{}_{a}\otimes(\gen{\tilde S}_{-1-m})^a{}_{\alpha}
- (\gen{S}_m)^a{}_{\alpha}\otimes(\gen{\tilde Q}_{-1-m})^\alpha{}_{a}
\nl\qquad
+ \gen{C}_m\otimes\gen{\tilde{B}}_{-1-m}
+ \gen{B}_m\otimes\gen{\tilde C}_{-1-m}
\Bigr].
\>
Formally, it looks similar to the standard $\alg{u}(2|2)$ classical r-matrix
in \eqref{eq:standardrasym}.
The only difference is that for the $\alg{su}(2)\times\alg{su}(2)$
generators there are additional terms
with inverted level numbers in the second line.

To recover the above fundamental r-matrix in \tabref{tab:rCoeff}
a representation quite different from the standard evaluation representation was used:
\<
(\gen{Q}_m)^\alpha{}_{b} \simeq
(\gen{\tilde Q}_m)^\alpha{}_{b} \earel{\simeq}
\gen{Q}^\alpha{}_{b}(x^m\oldPi\indup{b} + x^{-m}\oldPi\indup{f}),
\nln
(\gen{S}_m)^a{}_{\beta} \simeq
(\gen{\tilde S}_m)^a{}_{\beta} \earel{\simeq}
\gen{S}^a{}_{\beta}(x^m\oldPi\indup{f} + x^{-m}\oldPi\indup{b}),
\nln
(\gen{R}_m)^a{}_{b} \earel{\simeq}
\gen{R}^a{}_{b}\frac{x^{m+1} -  x^{-m-1}}{x-x^{-1}}\,,
\nln
(\gen{\tilde R}_m)^a{}_{b} \earel{\simeq}
-\gen{R}^a{}_{b}\frac{x^{m-1} -  x^{-m+1}}{x-x^{-1}}\,,
\nln
(\gen{L}_m)^\alpha{}_{\beta} \earel{\simeq}
\gen{L}^\alpha{}_{\beta}\frac{x^{m+1} -  x^{-m-1}}{x-x^{-1}}\,,
\nln
(\gen{\tilde L}_m)^\alpha{}_{\beta} \earel{\simeq}
-\gen{L}^\alpha{}_{\beta}\frac{x^{m-1} -  x^{-m+1}}{x-x^{-1}}\,,
\nln
\gen{C}_m \simeq \gen{\tilde C}_m \earel{\simeq}
\frac{1}{2}\,\frac{x^{m+1} +  x^{-m-1}}{x-x^{-1}}\,,
\nln
\gen{B}_m \simeq \gen{\tilde B}_m \earel{\simeq}
\frac{1}{2}\,(x^m - x^{-m})\,.
\>
This representation is quite unusual since it does not treat
the different generators on an equal footing
and since it makes a distinction between bosons and fermions
by means of the projection operators $\oldPi\indup{b,f}$.
The argument of the authors of \cite{Moriyama:2007jt}
not to use the standard evaluation representation
$\gen{J}_n = x^n\gen{J}$ was due to the fact that that
this would lead to a classical r-matrix with poles only in $x_1=x_2$
and none at $x_1 = 1/x_2$, in agreement with \tabref{tab:rCoeff}.
Furthermore, the proposed Lie brackets look quite complicated,
they are not of a standard loop algebra form,
and we shall not reproduce them here.
We do not know whether the brackets obey the Jacobi identities
and whether the r-matrix satisfies the CYBE using these brackets;
no statement was made in \cite{Moriyama:2007jt}.

In contrast, it was shown in \cite{Beisert:2007ds}
that the $\psucentral$ fundamental R-matrix
is invariant under Yangian generators
for an ordinary evaluation representation
with evaluation parameter $iu$ with $u=x+1/x$.
A reason for this superficial mismatch
was proposed in \cite{Moriyama:2007jt}:
The work \cite{Beisert:2007ds} was formulated in Drinfeld's
first realisation of the Yangian,
and the work \cite{Moriyama:2007jt} was formulated in the second.
In principle there might be a non-trivial map between the two realisations
which would make the two representations equivalent.

We should note that the procedure applied in \cite{Moriyama:2007jt}
is not necessarily unique.
There may be several representations leading to the
same fundamental r-matrix in \tabref{tab:rCoeff}
upon inserting into the above classical r-matrix
because often several terms of the classical r-matrix
contribute to a single matrix element of the fundamental r-matrix.
One needs to make a choice of how to distribute the individual terms
to these contributions from the classical r-matrix.
Apparently the choice of a symmetric distribution
was made in \cite{Moriyama:2007jt} which led to the above representation.
Furthermore, it was admitted in \cite{Moriyama:2007jt} that
the resulting algebra is not unique.

Here we note that a loop variable $iu$ with $u = x + 1/x$ automatically
leads to poles at $x_1=x_2$ and $x_1 = 1/x_2$ as required for the
fundamental r-matrix. Instead of producing just the right poles
in each term, one might in this way attempt to cancel the wrong poles.
As the determination of the representation is not unique, this can indeed
lead to the same fundamental r-matrix using the above (or a similar)
classical r-matrix.
In the remainder of this section we shall make an alternative proposal
for a normal evaluation representation based on $iu$,
a consistent Lie algebra and a classical r-matrix obeying the CYBE.
This leads to a direct analog of the Yangian considered in \cite{Beisert:2007ds}.
At this point we cannot say whether the proposal of \cite{Moriyama:2007jt}
is consistent with ours and merely represents a very different choice of basis.
A change of basis can indeed lead
to superficially quite different algebras
as the example in \secref{sec:algu22} shows.
In any case, we find our proposal more natural and it is probably
easier to deal with because it employs standard evaluation representations
and an (almost) standard loop algebra.

%%%%%%%%%%%%%%%%%%%%%%%%%%%%%%%%%%%%%%%%%%%%%%%%%%%%%%%%%%%%%%%%%%%%%%%%%%%%%%%%
\subsection{Observation}

The standard form for the classical r-matrix \eqref{eq:standardrasym}
makes use of the quadratic Casimir.
Unfortunately, it does not exist for our algebra $\alg{h}$
because the $\alg{sl}(2)$ automorphisms would be required
to complement the central charges, see \secref{sec:sl2auto}.
Nevertheless the quadratic Casimir operator for
$\alg{psu}(2|2)$ can be written within $\alg{h}$
\[\label{eq:psucasimir}
\mathcal{T}=
\half\gen{R}^c{}_d\gen{R}^d{}_c
-\half\gen{L}^\gamma{}_\delta\gen{L}^\delta{}_\gamma
+\half\gen{Q}^\gamma{}_d\gen{S}^d{}_\gamma
-\half\gen{S}^c{}_\delta\gen{Q}^\delta{}_c.
\]
The corresponding two-site operator reads
\[
\mathcal{T}_{12}=
\gen{R}^c{}_d\otimes\gen{R}^d{}_c
-\gen{L}^\gamma{}_\delta\otimes\gen{L}^\delta{}_\gamma
+\gen{Q}^\gamma{}_d\otimes\gen{S}^d{}_\gamma
-\gen{S}^c{}_\delta\otimes\gen{Q}^\delta{}_c.
\]

Here we make the crucial observation that all the
off-diagonal elements of the r-matrix in
\tabref{tab:rCoeff} are generated by the operator
$\mathcal{T}_{12}/(iu_1-iu_2)$.
To make this statement more transparent,
we have written the coefficients in \tabref{tab:rCoeff}
in an alternative form using the
coefficients $a,b,c,d$ \eqref{eq:abcd} which determine
the action of supercharges.
The diagonal elements, however,
are not reproduced correctly.
Nevertheless the remainder takes a
peculiar form in which two signs
only depend on whether the state the
r-matrix acts upon consists of bosons or fermions.
Formally, we can achieve full agreement with
the fundamental r-matrix by the following expression
\[\label{eq:classicalr}
r_{12}
=
\frac{\mathcal{T}_{12}
-\mathcal{T}\gen{D}^{-1}\otimes \gen{D}
-\gen{D}\otimes\mathcal{T}\gen{D}^{-1}
}{iu_1-iu_2}\,.
\]
However, this is not an element of $\gen{h}\otimes\gen{h}$ but rather
of its enveloping algebra.
Furthermore, the element $\gen{D}$ is strictly speaking not invertible.
That means that formally the expression \eqref{eq:classicalr}
may be used to compute the r-matrix in evaluation representations,
but it is not a universal r-matrix.
Before we continue, let us rewrite the r-matrix in a slightly
different manner
\<\label{eq:rstrongweird}
r_{12}
\eq
\frac{\mathcal{T}_{12}
-(iu_1/iu_2)\mathcal{T}\gen{C}^{-1}\otimes \gen{C}
-(iu_2/iu_1)\gen{C}\otimes\mathcal{T}\gen{C}^{-1}
}{iu_1-iu_2}
\nln
\eq
\frac{\mathcal{T}_{12}
-\mathcal{T}\gen{C}^{-1}\otimes \gen{C}
-\gen{C}\otimes\mathcal{T}\gen{C}^{-1}
}{iu_1-iu_2}
-\frac{\mathcal{T}\gen{C}^{-1}\otimes \gen{C}}{iu_2}
+\frac{\gen{C}\otimes\mathcal{T}\gen{C}^{-1}}{iu_1}\,.
\>
Clearly we have not gained anything by this transformation, but this
will be a more convenient starting point for our further analysis.

%%%%%%%%%%%%%%%%%%%%%%%%%%%%%%%%%%%%%%%%%%%%%%%%%%%%%%%%%%%%%%%%%%%%%%%%%%%%%%%%
\subsection{A Deformation of the $\alg{u}(2|2)$ Loop Algebra}
\label{sec:algdef}

To accommodate the r-matrix in a Lie bialgebra
it should consist of bilinear terms in the generators only.
Instead of $\mathcal{T}\gen{C}^{-1}$ we should have a single Lie generator $\gen{B}$.
Let us therefore examine the commutators of this combination
and see if we can interpret them as Lie brackets
\<
\comm{\mathcal{T}\gen{C}^{-1}}{\gen{Q}^\alpha{}_b}\eq
+\gen{Q}^\alpha{}_b
+i\alpha\varepsilon^{\alpha\gamma}\varepsilon_{bd}\gen{D}\gen{C}^{-1}\gen{S}^d{}_\gamma,
\nln
\comm{\mathcal{T}\gen{C}^{-1}}{\gen{S}^a{}_\beta}\eq
-\gen{S}^a{}_\beta
+i\alpha^{-1}\varepsilon^{ac}\varepsilon_{\beta\delta}\gen{D}\gen{C}^{-1}\gen{Q}^\delta{}_c.
\>
The resulting linear terms are clearly okay. For the cubic terms
we note that $u=2CD^{-1}$, which means that
we may interpret the combination $\gen{D}\gen{C}^{-1}$
as a shift in the level of a loop algebra generator.
If we introduce $\gen{B}$ such that its brackets
coincide with commutators of $\mathcal{T}\gen{C}^{-1}$,
the loop algebra becomes
\<\label{eq:BQS}
\comm{\gen{B}_m}{(\gen{Q}_n){}^\alpha{}_b}\eq
+(\gen{Q}_{m+n}){}^\alpha{}_b
-2\alpha\beta\varepsilon^{\alpha\gamma}\varepsilon_{bd}(\gen{S}_{m+n-1}){}^d{}_\gamma,
\nln
\comm{\gen{B}_m}{(\gen{S}_n){}^a{}_\beta}\eq
-(\gen{S}_{m+n}){}^a{}_\beta
-2\alpha^{-1}\beta\varepsilon^{ac}\varepsilon_{\beta\delta}(\gen{Q}_{m+n-1}){}^\delta{}_c.
\>
Here we have introduced a parameter $\beta$ which
in our case equals $\beta=1$.%
\footnote{In fact one may keep
the parameter $\beta$ arbitrary
if one inserts it into the relations \eqref{eq:PKident} as well.
Essentially $\beta$ corresponds to a rescaling of $g$.}
In fact these relations
are very reminiscent of the automorphism in $\alg{u}(2|2)$.
The deformation parameter $\beta$ in fact interpolates between the
standard $\alg{u}(2|2)$ loop algebra, which we get for $\beta=0$,
and our algebra.%
\footnote{Note that we also have $\alg{u}(2|2)$ symmetry in the
alternative limit discussed in \protect\secref{sec:weak},
where we provide further details.}
The remaining brackets of $\gen{B}$ should be trivial
\[\label{eq:Brest}
\comm{\gen{B}_m}{\gen{C}_n}=
\comm{\gen{B}_m}{(\gen{R}_n)^a{}_b}=
\comm{\gen{B}_m}{(\gen{L}_n)^\alpha{}_\beta}=0.
\]
Finally, because we have $u=2CD^{-1}$,
we may identify $\gen{D}_n=2i\gen{C}_{n-1}$ and obtain for the
brackets of supercharges
\<\label{eq:QSbi}
\acomm{(\gen{Q}_m){}^\alpha{}_b}{(\gen{S}_n){}^c{}_\delta}\eq
\delta^c_b(\gen{L}_{m+n}){}^\alpha{}_\delta
+\delta^\alpha_\delta(\gen{R}_{m+n}){}^c{}_b
+\delta^c_b\delta^\alpha_\delta (\gen{C}_{m+n}){},
\nln
\acomm{(\gen{Q}_m){}^{\alpha}{}_{b}}{(\gen{Q}_n){}^{\gamma}{}_{d}}\eq
2\alpha\beta\varepsilon^{\alpha\gamma}\varepsilon_{bd}\gen{C}_{m+n-1},
\nln
\acomm{(\gen{S}_m){}^{a}{}_{\beta}}{(\gen{S}_n){}^{c}{}_{\delta}}\eq
-2\alpha^{-1}\beta\varepsilon^{ac}\varepsilon_{\beta\delta}\gen{C}_{m+n-1}.
\>
The brackets of $\alg{su}(2)\times\alg{su}(2)$
are undeformed and given in \eqref{eq:su2su2}
(supplemented with additive levels of the loop algebra).
It is not difficult to confirm that
these brackets obey the Jacobi identity
for arbitrary $\alpha,\beta$
and therefore they define a family of Lie algebras.
In fact, the algebra can be embedded into 
the regular $\alg{u}(2|2)$ loop algebra 
as we shall see below in \secref{sec:algu22}.

The action of $\gen{B}$ on the fundamental representation
for $\beta=1$ should be equal to $\mathcal{T}\gen{C}^{-1}$
which yields
\[\label{eq:Bclassfund}
\gen{B}\state{\phi^a}=-\frac{1}{4C}\,\state{\phi^a},
\qquad
\gen{B}\state{\psi^\alpha}=+\frac{1}{4C}\,\state{\psi^\alpha}.
\]
%

%%%%%%%%%%%%%%%%%%%%%%%%%%%%%%%%%%%%%%%%%%%%%%%%%%%%%%%%%%%%%%%%%%%%%%%%%%%%%%%%
\subsection{Classical r-matrix and Cobrackets}

We can now write down a classical r-matrix for our Lie algebra
by substituting $\gen{B}$ for $\mathcal{T}\gen{C}^{-1}$ in \eqref{eq:rstrongweird}
\[\label{eq:rstrong}
r_{12}
=
\frac{\mathcal{T}_{12}
-\gen{B}\otimes \gen{C}
-\gen{C}\otimes \gen{B}
}{iu_1-iu_2}
-\frac{\gen{B}\otimes \gen{C}}{iu_2}
+\frac{\gen{C}\otimes \gen{B}}{iu_1}\,.
\]
This expression assumes evaluation representations, but we
can reexpress it in full generality using loop algebra generators
\[\label{eq:rclass}
r
=r\indup{\alg{psu}(2|2)}-
\sum_{m=-1}^\infty
\gen{B}_{-1-m}\otimes\gen{C}_{m}
-
\sum_{m=+1}^\infty
\gen{C}_{-1-m}\otimes\gen{B}_{m}
\]
with the classical r-matrix $r\indup{\alg{psu}(2|2)}$
for $\alg{psu}(2|2)$
\<\label{eq:rclasspsu}
r\indup{\alg{psu}(2|2)}
\eq
+
\sum_{m=0}^\infty
(\gen{R}_{-1-m})^c{}_d\otimes(\gen{R}_{m})^d{}_c
-
\sum_{m=0}^\infty
(\gen{L}_{-1-m})^\gamma{}_\delta\otimes(\gen{L}_{m})^\delta{}_\gamma
\nl
+
\sum_{m=0}^\infty
(\gen{Q}_{-1-m})^\gamma{}_d\otimes(\gen{S}_{m})^d{}_\gamma
-
\sum_{m=0}^\infty
(\gen{S}_{-1-m})^c{}_\delta\otimes(\gen{Q}_{m})^\delta{}_c.
\>

This expression is almost the standard form for $\alg{u}(2|2)[iu,(iu)^{-1}]$,
but note that the lower bound on the sum for the $\gen{B}$-$\gen{C}$ terms
is shifted by $\pm1$ due to the extra terms in \eqref{eq:rstrong}.
To motivate the extra term $\gen{C}_{-1}\wedge\gen{B}_{0}$
recall that the coproduct \eqref{eq:braid}
is not the ordinary Yangian coproduct
but contains an additional braiding factor.
For \emph{undeformed} $\alg{u}(2|2)[iu,(iu)^{-1}]$ this braided coproduct
can easily be obtained from the standard coproduct
via a Reshetikhin twist transformation \cite{Reshetikhin:1990ep}
\[
\copro(\gen{J})=\mathcal{F} \copro_0(\gen{J})\mathcal{F}^{-1},
\qquad
\rmat=\perm(\mathcal{F})\rmat_0\mathcal{F}^{-1}.
\]

with $\mathcal{F}=\exp(-g^{-1}\gen{C}_{-1}\otimes \gen{B}_0)$.
The requirements for the transformation in \cite{Reshetikhin:1990ep}
are satisfied because the coproducts of the Cartan elements $\gen{C}_{-1}$
and $\gen{B}_{0}$ are both trivial.
The twist $\mathcal{F}=1\otimes 1+g^{-1}f+\order{g^{-2}}$
will contribute to the classical r-matrix by the term $-f+\perm(f)$.
Indeed, $\gen{C}_{-1}\wedge\gen{B}_{0} =-f+\perm(f)$
is the classical contribution from the twist.

It is also straightforward to include the AFS phase \eqref{eq:AFSphaseshift}
by adding to \eqref{eq:rclass}
\[
r_0=-\gen{C}_{-1}\wedge\gen{C}_0.
\]
Note that curiously one can combine the extra term discussed above with the phase
into $\gen{C}_{-1}\wedge(\gen{B}_0-\gen{C}_0)$.
This shift clearly has no impact on any of the relevant properties of
classical r-matrices because the $\gen{C}_n$ are central elements of the algebra.
In fact, one can incorporate arbitrary phase because
terms of the sort $\gen{C}_m\wedge\gen{C}_n$ do
not modify any of the relevant properties
of classical r-matrices.%
\footnote{In a lift to the quantum theory, however, the
Hopf coproduct of central charges is not necessarily trivial
and therefore the set of possible twists 
will be reduced by demanding quasi-triangularity.}
With these terms one can represent
an arbitrary antisymmetric function
of the two variables $u_1$ and $u_2$
by
\[
r_0=\sum_{m,n=-\infty}^\infty c_{m,n} \gen{C}_m\wedge\gen{C}_n
\]
with antisymmetric coefficients $c_{m,n}$.
Note that these contributions can also be viewed as a
Reshetikhin twist similarly to the above discussion.

We can now determine the cobrackets from
the r-matrix via the standard relation \eqref{eq:classcoco};
the results are summarised in \tabref{tab:cobra}.
These expressions agree exactly with the expected cobrackets
for centrally extended $\alg{psu}(2|2)$ in \eqref{eq:cobraextsu22}
when we set the deformation parameter $\beta=1$.
We also see that cobracket
$\coprocl(\gen{B}_1)=\gen{Q}^\alpha{}_b\wedge\gen{S}^b{}_\alpha$,
which is not part of centrally extended $\alg{su}(2|2)$,
is consistent with the coproduct of the combination $2i\mathcal{T}\gen{D}^{-1}$
in the Hopf algebra.

\begin{table}\centering
\<
\coprocl(\gen{C}_n)\eq 0
\nln
\coprocl(\gen{B}_n)\eq
+\sum_{k=0}^{n-1}
(\gen{Q}_k)^\alpha{}_b\wedge(\gen{S}_{n-1-k})^b{}_\alpha
\nl
+\sum_{k=1}^{n-1}
\alpha^{-1}\beta\varepsilon^{bd}\varepsilon_{\alpha\gamma}
(\gen{Q}_{k-1})^\alpha{}_b\wedge(\gen{Q}_{n-1-k})^\gamma{}_d
\nl
-\sum_{k=1}^{n-1}
\alpha\beta\varepsilon^{\beta\delta}\varepsilon_{ac}
(\gen{S}_{k-1})^a{}_\beta\wedge(\gen{S}_{n-1-k})^c{}_\delta
\nln
\coprocl(\gen{R}_n)^a{}_b\eq
+\sum_{k=0}^{n-1}
(\gen{R}_k)^a{}_c\wedge(\gen{R}_{n-1-k})^c{}_b
\nl
-\sum_{k=0}^{n-1}
\Bigl[
(\gen{S}_k)^a{}_\gamma\wedge(\gen{Q}_{n-1-k})^\gamma{}_b
-\half\delta^a_b\,(\gen{S}_k)^d{}_\gamma\wedge(\gen{Q}_{n-1-k})^\gamma{}_d
\Bigr]
\nln
\coprocl(\gen{L}_n)^\alpha{}_\beta\eq
-\sum_{k=0}^{n-1}
(\gen{L}_k)^\alpha{}_\gamma\wedge(\gen{L}_{n-1-k})^\gamma{}_\beta
\nl
+\sum_{k=0}^{n-1}
\Bigl[
(\gen{Q}_k)^\alpha{}_c\wedge(\gen{S}_{n-1-k})^c{}_\beta
-\half\delta^\alpha_\beta\,(\gen{Q}_k)^\delta{}_c\wedge(\gen{S}_{n-1-k})^c{}_\delta
\Bigr]
\nln
\coprocl(\gen{Q}_{n})^\alpha{}_b\eq
-\sum_{k=0}^{n-1}
(\gen{L}_k)^\alpha{}_\gamma\wedge(\gen{Q}_{n-1-k})^\gamma{}_b
-\sum_{k=0}^{n-1}
(\gen{R}_k)^c{}_b\wedge(\gen{Q}_{n-1-k})^\alpha{}_c
\nl
-\sum_{k=0}^{n}
\gen{C}_{k-1}\wedge(\gen{Q}_{n-k})^\alpha{}_b
+\sum_{k=0}^{n-1}
2\alpha\beta\varepsilon^{\alpha\gamma}\varepsilon_{bd}\gen{C}_{k-1}\wedge(\gen{S}_{n-1-k})^d{}_\gamma
\nl
\nln
\coprocl(\gen{S}_n)^a{}_\beta\eq
+\sum_{k=0}^{n-1}
(\gen{R}_k)^a{}_c\wedge(\gen{S}_{n-1-k})^c{}_\beta
+\sum_{k=0}^{n-1}
(\gen{L}_k)^\gamma{}_\beta\wedge(\gen{S}_{n-1-k})^a{}_\gamma
\nl
+\sum_{k=0}^{n}
\gen{C}_{k-1}\wedge(\gen{S}_{n-k})^a{}_\beta
+\sum_{k=0}^{n-1}
2\alpha^{-1}\beta\varepsilon^{ac}\varepsilon_{\beta\delta}\gen{C}_{k-1}\wedge(\gen{Q}_{n-1-k})^\delta{}_c
\nn
\>
\caption{Cobrackets of the Lie bialgebra generators.}
\label{tab:cobra}
\end{table}

\medskip

Finally, we should prove the CYBE $\cybe{r}{r}=0$.
A convenient method is to split up the computation into three parts.
In the first part we shall set $\beta=0$ and also adjust all
lower bounds of the sums in \eqref{eq:rclass} to $m=0$.
Then we have the standard rational r-matrix for the
algebra $\grp{u}(2|2)[iu,(iu)^{-1}]$ which is known to satisfy the CYBE.

Secondly, we have omitted the term $\gen{C}_{-1}\wedge\gen{B}_{0}$
by adjusting the summation bounds.
As discussed above, this term originates from a Reshetikhin twist
and thus preserves the CYBE when $\beta=0$.
More explicitly, we add a term of the sort
\[\label{eq:gentwist}
r'=r+\gen{J}\wedge \gen{J}'.
\]
This changes the commutators in the CYBE to
\[\label{eq:gentwistcybe}
\cybe{r'}{r'}=
\cybe{r}{r}
+\gen{J}\mathbin{\wedge\wedge}\comm{\gen{J}'}{r}
-\gen{J}'\mathbin{\wedge\wedge}\comm{\gen{J}}{r}
+\gen{J}\wedge\gen{J}'\wedge\comm{\gen{J}}{\gen{J}'},
\]
where we define the double-wedge as
\[
\gen{J}^A\mathbin{\wedge\wedge}(\gen{J}^B\otimes\gen{J}^C)
:=
\gen{J}^A\otimes\gen{J}^B\otimes\gen{J}^C
-\gen{J}^B\otimes\gen{J}^A\otimes\gen{J}^C
+\gen{J}^B\otimes\gen{J}^C\otimes\gen{J}^A.
\]
For $\gen{J}=\gen{C}_{-1}$ and $\gen{J}'=\gen{B}_0$
all the additional terms vanish because
$\gen{C}_{-1}$ and $\gen{B}_0$ are both Cartan elements
and thus obey $\comm{\gen{B}_0}{r}=\comm{\gen{C}_{-1}}{r}=\comm{\gen{C}_{-1}}{\gen{B}_0}=0$.

It remains to confirm the CYBE for all
contributions proportional to the deformation $\beta$.
These originate from the brackets
$\comm{\gen{B}}{\gen{Q}}$,
$\comm{\gen{B}}{\gen{S}}$, as well as
$\acomm{\gen{Q}}{\gen{Q}}$ and
$\acomm{\gen{S}}{\gen{S}}$.
It is relatively easy to confirm that these terms cancel.
Here the modification of the summation bounds in \eqref{eq:rclass}
is crucial; without it some terms would remain.%
\footnote{This is in agreement with the fact that
$\gen{C}_{-1}\wedge\gen{B}_{0}$ does \emph{not}
correspond to a Reshetikhin twist
of the \emph{deformed} algebra because $\comm{\gen{B}_0}{r}=\delta(\gen{B}_0)\neq 0$
and thus $\cybe{r'}{r'}\neq0$ according to \protect\eqref{eq:gentwistcybe}.}

%%%%%%%%%%%%%%%%%%%%%%%%%%%%%%%%%%%%%%%%
\subsection{The Deformed Loop Algebra as a Classical Double}

In this section we want to show that the deformed loop algebra
including the r-matrix \eqref{eq:classicalr}
can be obtained via a classical double construction.
If a bialgebra can be written as a double it is automatically quasi-triangular.
In general, the classical double $\grp{D}(\alg{g})$
of a Lie bialgebra $\alg{g}$ is the vector space
\[
\grp{D}(\alg{g}) = \alg{g}\oplus\alg{g}^*.
\]
The Lie brackets of $\grp{D}(\alg{g})$ read
\[
\comm{\gen{J}^A}{\gen{J}^B}=F^{AB}_C\gen{J}^{C},\qquad
\comm{\gen{J}_A}{\gen{J}_B}=\tilde{F}_{AB}^C\gen{J}_{C},\qquad
\comm{\gen{J}^A}{\gen{J}_B}= \tilde F^{A}_{BC}\gen{J}^{C} - F^{AC}_{B}\gen{J}_{C},
\]
with $\gen{J}^A$ forming a basis of $\alg{g}$ and $\gen{J}_A$
being the corresponding dual basis of $\alg{g}^*$.
The coalgebra structure of $\grp{D}(\alg{g})$
is simply given by the canonical r-matrix
\[\label{eq:doubler}
\crmat = \gen{J}^A\otimes\gen{J}_A.
\]
One can convince oneself that the induced cobracket reads
\<
\coprocl(\gen{J}^A)= \tilde{F}^{A}_{BC}\gen{J}^{B}\otimes\gen{J}^{C},\qquad
\coprocl(\gen{J}_A)= -F_{A}^{BC}\gen{J}_{B}\otimes\gen{J}_{C}.
\>
and that the bialgebra $\grp{D}(\alg{g})$ is quasi-triangular.

In the standard loop algebra $\alg{g}[u,u^{-1}]$ of a Lie algebra $\alg{g}$
with non-degenerate invariant bilinear form $C_{AB}$ we may take the decomposition
into the subalgebra $\alg{g}^+ = \alg{g}[u]$
consisting of generators with non-negative powers in $u$
and the subalgebra $\alg{g}^- = u^{-1}\alg{g}[u^{-1}]$
consisting of generators of negative powers in $u$.
Then we indeed have%
\footnote{Strictly speaking, we should pair polynomials $\alg{g}[u]$
with formal power series $u^{-1}\alg{g}[[u^{-1}]]$,
resulting in the double $\alg{g}((u^{-1}))$ being the field of fractions
of $u^{-1}\alg{g}[[u^{-1}]]$.
We will ignore these mathematical subtleties
and always allow for infinite power series,
implicitly assuming that we are
working in some suitable topological completions of the considered algebras.
For a more mathematical treatment
we refer the reader to e.g.\ \cite{Chari:1994pz}.}
$(\alg{g}^+)^\ast=\alg{g}^-$ and
$\grp{D}(\alg{g}^+)  = \alg{g}^+ \oplus \alg{g}^- = \alg{g}[u,u^{-1}]$.
A dual pairing between the two subalgebras is given by
\[
(\gen{J}^A_n,\gen{J}^B_m) = -\delta_{n,-m-1}C^{AB}
\]
with $C^{AB}$ the Cartan--Killing matrix of $\alg{g}$.
It defines a consistent cobracket on $\alg{g}^+$ from the
bracket of $\alg{g}^-$.
The induced classical r-matrix in \eqref{eq:doubler} then reads
\[
\crmat = -\sum_{n=0}^\infty C_{AB}\gen{J}^A_n\otimes\gen{J}^B_{-n-1},
\]
which is precisely one of the asymmetric r-matrices of \eqref{eq:standardrasym}.
The other one is obtained by exchanging $\alg{g}^+$ and $\alg{g}^-$,
i.e.\ considering the double of $\alg{g}^-$.

The case of our deformed $\alg{u}(2|2)$
works almost in the same way as for generic loop algebras.
Due to the deformed commutation relations of the automorphisms $\gen{B}_n$
and the identification of the loop variable
with the central charges we should actually set
\<
\alg{g}^+\eq\langle \gen{R}_n, \gen{L}_n, \gen{Q}_n, \gen{S}_n, \gen{C}_{n-1}, \gen{B}_{n+1}\rangle_{n\geq 0},
\nln
\alg{g}^-\eq\langle \gen{R}_{-1-n}, \gen{L}_{-1-n}, \gen{Q}_{-1-n}, \gen{S}_{-1-n}, \gen{C}_{-2-n}, \gen{B}_{-n}\rangle_{n\geq 0}.
\>
With this assignment $\alg{g}^+$ and $\alg{g}^-$
are indeed sub-bialgebras of $\alg{g}$,
i.e.\ the brackets \eqref{eq:BQS,eq:QSbi} and the cobrackets (\tabref{tab:cobra})
close on $\alg{g}^+$ and $\alg{g}^-$, respectively;
the level shift for $\gen{B}_n$ and $\gen{C}_n$
is crucial for this property.
Furthermore, the r-matrix
\eqref{eq:rclass,eq:rclasspsu}
corresponds to a dual pairing between $\alg{g}^+$ and $\alg{g}^-$.
These properties suffice to show that
the deformed $\alg{u}(2|2)[u,u^{-1}]$ loop algebra
is the double $\grp{D}(\alg{g}^-)$.

%%%%%%%%%%%%%%%%%%%%%%%%%%%%%%%%%%%%%%%%%%%%%%%%%%%%%%%%%%%%%%%%%%%%%%%%%%%%%%%%
%%%%%%%%%%%%%%%%%%%%%%%%%%%%%%%%%%%%%%%%%%%%%%%%%%%%%%%%%%%%%%%%%%%%%%%%%%%%%%%%
\section{Relations to Standard Algebras}
\label{sec:relations}

In this section we relate the Lie bialgebra found in the
previous section to standard Lie (bi)algebras.
In particular we show that the Lie algebra is locally
(in the space of spectral parameters) 
isomorphic to the loop algebra of $\alg{u}(2|2)$,
but the coalgebra takes a non-standard form.
Furthermore we show how the complete bialgebra can be
obtained from a reduction of the loop algebra based
on the maximal extension $\alg{h}_+$ of $\alg{psu}(2|2)$.

%%%%%%%%%%%%%%%%%%%%%%%%%%%%%%%%%%%%%%%%%%%%%%%%%%%%%%%%%%%%%%%%%%%%%%%%%%%%%%%%
\subsection{Embedding into the Twisted $\alg{u}(2|2)$ Loop Algebra}
\label{sec:algu22}

In the following we shall try to express the algebra discussed in
\secref{sec:algdef} through elements of the loop algebra
of $\alg{u}(2|2)[iu,(iu)^{-1}]$
which we shall expand as Laurent polynomials
\[\label{eq:leveldef}
\bar{\gen{J}}^A_n:=(iu)^{n} \bar{\gen{J}}^A.
\]
%

%%%%%%%%%%%%%%%%%%%%%%%%%%%%%%%%%%%%%%%%
\paragraph{Standard Loop Algebra.}

First of all, it is reasonable to expect that
the $\alg{su}(2)\times\alg{su}(2)$ generators are not deformed
\[
\gen{R}^a{}_b= \bar{\gen{R}}^a{}_b,\quad
\gen{L}^\alpha{}_\beta = \bar{\gen{L}}^\alpha{}_\beta.
\]
For the $\alg{u}(1)\times\alg{u}(1)$ generators $\gen{B},\gen{C}$ acting as
$\comm{\bar{\gen{B}}}{\bar{\gen{J}}^A}=[A]\gen{J}^A$ and
$\comm{\bar{\gen{J}}^A}{\bar{\gen{C}}}=0$ we allow for a rescaling
\[
\gen{C}= e \bar{\gen{C}},\quad
\gen{B}= f \bar{\gen{B}}.
\]
Finally we make the ansatz that the fermionic generators can be mixed by a general $2\times 2$ matrix
\<
\gen{Q}^\alpha{}_b\eq
a \bar{\gen{Q}}^\alpha{}_b
+b\varepsilon^{\alpha\gamma}\varepsilon_{bd}\bar{\gen{S}}^d{}_\gamma,
\nln
\gen{S}^a{}_\beta\eq d \bar{\gen{S}}^a{}_\beta
+c\varepsilon^{ac}\varepsilon_{\beta\delta}\bar{\gen{Q}}^\delta{}_c.
\>
Note that the coefficients $a,\ldots,f$ are assumed to be functions of $iu$.
A generator $\bar{\gen{J}}$ multiplied by a function of $iu$
is understood as a generator of the loop algebra according
to the identification \eqref{eq:leveldef}.
Thus the new generators
will generically be linear combinations of
generators at different levels of the loop algebra.
The latter two redefinitions are obviously consistent with
the undeformed $\alg{su}(2)\times\alg{su}(2)$ transformation rules.

Consider now the brackets of supercharges. For example
the bracket
\[
\bigacomm{(\gen{Q}_n)^\alpha{}_b}{(\gen{Q}_m)^\gamma{}_d}=
2\alpha\beta \varepsilon^{\alpha\gamma}\varepsilon_{bd}\gen{C}_{m+n-1}
\]
after substitution and evaluation of $\alg{u}(2|2)$ brackets reads
\[
2ab \varepsilon^{\alpha\gamma}\varepsilon_{bd}(iu)^{m+n}\bar{\gen{C}} =
2\alpha\beta \varepsilon^{\alpha\gamma}\varepsilon_{bd} (iu)^{m+n-1}e\bar{\gen{C}},
\]
or $ab=\alpha\beta e/iu$ for short.
The brackets of supercharges thus lead to the following four constraints
\[\label{eq:abcde}
ad-bc=1, \quad
ad+bc=e, \quad
ab=\alpha\beta e/iu,\quad
cd=-\alpha^{-1}\beta e/iu,
\]
which have two solutions for $b,c,d,e$
in terms of $\alpha,\beta,iu$.
Furthermore we should consider the brackets of $\gen{B}$, for example
\[
\comm{\gen{B}}{\gen{Q}^\alpha{}_b}
=\gen{Q}^\alpha{}_b
-2\alpha\beta\varepsilon^{\alpha\gamma}\varepsilon_{bd} (iu)^{-1}\gen{S}^d{}_\gamma.
\]
After substitution and evaluation this leads to the relation
\[
fa \bar{\gen{Q}}^\alpha{}_b
-fb\varepsilon^{\alpha\gamma}\varepsilon_{bd}\bar{\gen{S}}^d{}_\gamma
=
a \bar{\gen{Q}}^\alpha{}_b
+b\varepsilon^{\alpha\gamma}\varepsilon_{bd}\bar{\gen{S}}^d{}_\gamma
-2\alpha\beta\varepsilon^{\alpha\gamma}\varepsilon_{bd} d(iu)^{-1} \bar{\gen{S}}^d{}_\gamma
-2\alpha\beta c(iu)^{-1}\bar{\gen{Q}}^\alpha{}_b.
\]
We can write this, together with the bracket $\comm{\gen{B}}{\gen{S}}$ as
\[\begin{array}{rclcrcl}
fa\eq a-2\alpha\beta c/iu,&\quad &
-fb\eq b-2\alpha\beta d/iu,\\[3pt]
fc\eq -c-2\alpha^{-1}\beta a/iu,&\quad&
-fd\eq -d-2\alpha^{-1}\beta b/iu.
\end{array}
\]
All these equations are equivalent to $fe=1$
upon imposing \eqref{eq:abcde}.

Altogether we find the solution
\[
f=1/e=\sqrt{1-\frac{4\beta^2}{u^2}}\,,\quad
ad=\frac{1+e}{2}\,,\quad
b=\frac{\alpha\beta e}{iua}\,,\quad
c=-\frac{\alpha^{-1}\beta e}{iud}\,.
\]
The value of $a$ (or $d$) is not fixed;
a convenient choice is given by
\[
a=\bar\alpha\sqrt{\frac{1+e}{2}}\,,\qquad
d=\bar\alpha^{-1}\sqrt{\frac{1+e}{2}}\,,
\]
in which case the $2\times 2$ matrix defined by the four elements $a,b,c,d$ becomes
quasi-orthogonal (and for $\alpha=\bar\alpha=1$ strictly orthogonal).

The solution shows that the algebra of \secref{sec:algdef}
can be embedded into the algebra
of functions $\Complex/\set{0}\to\alg{u}(2|2)$
with Lie brackets canonically defined as for loop algebras.
Note however, that the functions $a,\ldots,f$
are not meromorphic on $\bar\Complex$
and not holomorphic on $\Complex/\set{0}$
as required for an embedding
into the loop algebra $\alg{u}(2|2)[iu,(iu)^{-1}]$.
Expanding the square roots at $u=0$ or $u=\infty$ leads
to Laurent series over the levels \eqref{eq:leveldef}.
After the expansion the singularities of the square roots
cannot be seen, and thus the proposed change of basis
works only locally in the complex spectral parameter plane.
As we shall see shortly, the global properties are changed.

Despite these problems, the above transformation
can at least be understood as a way to show
that the algebra in \secref{sec:algdef} satisfies Jacobi identities
because $\alg{u}(2|2)$ does.
One might work with the above $\alg{u}(2|2)$-manifest basis,
but in the bialgebra this would lead to a
rather complicated r-matrix.
In the basis of \secref{sec:algdef}
the r-matrix takes almost the same form as for $\alg{u}(2|2)[iu,(iu)^{-1}]$,
but at the cost of slightly deformed Lie brackets.

%%%%%%%%%%%%%%%%%%%%%%%%%%%%%%%%%%%%%%%%
\paragraph{Twisted Loop Algebra.}

In order to understand the global structure of the spectral parameter plane
let us introduce a transformation that removes the
square root singularities
\[
z^4=\frac{u+2\beta}{u-2\beta}\,,\qquad
u=2\beta\,\frac{z^4+1}{z^4-1}\,.
\]
Using the spectral parameter $z$, the generators
of the deformed loop algebra in \secref{sec:algdef}
can be expressed through \emph{meromorphic} functions $\bar\Complex\to\alg{u}(2|2)$
as follows
\<\label{eq:embed}
\gen{C}_n\eq(2i\beta)^n\, \lrbrk{\frac{z^4+1}{z^4-1}}^n\, \frac{z^4+1}{2z^2}\,\bar{\gen{C}},
\nln
\gen{B}_n\eq(2i\beta)^n\,\lrbrk{\frac{z^4+1}{z^4-1}}^n\, \frac{2z^2}{z^4+1}\,\bar{\gen{B}},
\nln
(\gen{Q}_n)^\alpha{}_b\eq
(2i\beta)^n\, \lrbrk{\frac{z^4+1}{z^4-1}}^n\,
\varepsilon_{bc}
\lrbrk{+\half z\,\bar{\gen{Q}}^{c \alpha}_+ +\half z^{-1}\bar{\gen{Q}}^{c \alpha}_-},
\nln
(\gen{S}_n)^a{}_\beta\eq
(2i\beta)^n\, \lrbrk{\frac{z^4+1}{z^4-1}}^n\,
i\alpha^{-1}\varepsilon_{\beta\gamma}
\lrbrk{-\half z\,\bar{\gen{Q}}^{a \gamma}_+ +\half z^{-1}\bar{\gen{Q}}^{a \gamma}_-}
\>
as well as
$\gen{R}_n= (iu)^n \bar{\gen{R}}$,
$\gen{L}_n= (iu)^n \bar{\gen{L}}$,
and where $\bar{\gen{Q}}^{a \beta}_\pm$ are the linear combinations
\[
\bar{\gen{Q}}^{a \beta}_\pm
=
-\bar\alpha\varepsilon^{ac}\bar{\gen{Q}}^\beta{}_c
\mp i\alpha\bar\alpha^{-1}\,\varepsilon^{\beta\gamma}\bar{\gen{S}}^a{}_\gamma.
\]

Curiously, the embedding is into the $\Integers_4$-invariant
part of the $\Integers_4$-automorphism of $\bar\Complex\to\alg{u}(2|2)$
defined by the gradings
\[
[\bar{\gen{R}}]=[\bar{\gen{L}}]\equiv 0,
\qquad
[\bar{\gen{Q}}_\pm]\equiv\pm 1,
\qquad
[\bar{\gen{B}}]=[\bar{\gen{C}}]\equiv2,
\qquad
[z]=-1.
\]
Furthermore the singularities of \eqref{eq:embed} in the $z$-plane are
very restrictive: There are poles of arbitrary degree at
eighth roots of unity. In addition, the generators $\bar{\gen{Q}}_{\pm}$
admit single poles at $z=\infty,0$, respectively,
and $\bar{\gen{C}}$ admits double poles at $z=\infty$ and $z=0$.
In other words, the deformed algebra in \secref{sec:algdef} can be embedded
into the $\Integers_4$-twisted algebra $\alg{u}(2|2)[z,z^{-1},(z-e^{\pi i\Integers/4})^{-1}]/\Integers_4$.
If one furthermore allows at most double poles at $z=0,\infty$ the algebras become
isomorphic.

%%%%%%%%%%%%%%%%%%%%%%%%%%%%%%%%%%%%%%%%%%%%%%%%%%%%%%%%%%%%%%%%%%%%%%%%%%%%%%%%
\subsection{Maximally Extended Algebra $\alg{h}_+$}
\label{sec:sl2auto}

We now show that the complete Lie bialgebra can be obtained as
a reduction of the standard loop algebra of the maximal extension $\alg{h}_+$
of $\alg{psu}(2|2)$.

%%%%%%%%%%%%%%%%%%%%%%%%%%%%%%%%%%%%%%%%
\paragraph{Loop Bialgebra.}

The maximal central extension $\alg{h}$ of $\alg{psu}(2|2)$ can be adjoined
by its external $\alg{sl}(2)$ automorphism \cite{Serganova:1985aa,Beisert:2006qh}
\[
\alg{h}_+ = \alg{sl}(2)\ltimes\alg{psu}(2|2)\ltimes\Reals^3.
\]
The automorphisms $\gen{B}^{\mathfrak{a}}{}_{\mathfrak{b}}$ obey the brackets
\<
\comm{\gen{B}^{\mathfrak{a}}{}_{\mathfrak{b}}}{\gen{B}^{\mathfrak{c}}{}_{\mathfrak{d}}}
\eq \delta_{\mathfrak{b}}^{\mathfrak{c}}\gen{B}^{\mathfrak{a}}{}_{\mathfrak{d}}
   -\delta^{\mathfrak{a}}_{\mathfrak{d}}\gen{B}^{\mathfrak{c}}{}_{\mathfrak{b}},
\nln
\comm{\gen{B}^{\mathfrak{a}}{}_{\mathfrak{b}}}{\gen{Q}^{c\delta\mathfrak{e}}}
\eq \delta_{\mathfrak{b}}^{\mathfrak{e}}\gen{Q}^{c\delta\mathfrak{a}}
   -\half\delta_{\mathfrak{b}}^{\mathfrak{a}}\gen{Q}^{c\delta\mathfrak{e}},
\nln
\comm{\gen{B}^{\mathfrak{a}}{}_{\mathfrak{b}}}{\gen{C}^{\mathfrak{c}}{}_{\mathfrak{d}}}
\eq \delta_{\mathfrak{b}}^{\mathfrak{c}}\gen{C}^{\mathfrak{a}}{}_{\mathfrak{d}}
   -\delta^{\mathfrak{a}}_{\mathfrak{d}}\gen{C}^{\mathfrak{c}}{}_{\mathfrak{b}},
\>
where we combined the supercharges $\alg{Q}^\alpha{}_b,\alg{S}^a{}_\beta$
into one doublet of generators $\gen{Q}^{a\beta \mathfrak{c}}$,
\[
\gen{Q}^{a\beta 1}=\varepsilon^{ac}\gen{Q}^\beta{}_c,
\quad
\gen{Q}^{a\beta 2}=\varepsilon^{\beta\gamma}\gen{S}^a{}_\gamma,
\]
and the charges $\gen{C},\gen{P},\gen{K}$
into one triplet $\gen{C}^{\mathfrak{a}}{}_{\mathfrak{b}}$ with
\[
\gen{C}^1{}_1=-\gen{C}^2{}_2=\gen{C},
\quad
\gen{C}^1{}_2=\gen{P},
\quad
\gen{C}^2{}_1=-\gen{K}.
\]
Consequently the bracket of combined supercharges reads
\[\acomm{\gen{Q}^{a\beta\mathfrak{c}}}{\gen{Q}^{d\epsilon\mathfrak{f}}}
=
-\varepsilon^{\beta\epsilon}\varepsilon^{\mathfrak{cf}}\varepsilon^{ak}\gen{R}^d{}_k
+\varepsilon^{ad}\varepsilon^{\mathfrak{cf}}\varepsilon^{\beta\kappa}\gen{L}^\epsilon{}_\kappa
+\varepsilon^{ad}\varepsilon^{\beta\epsilon}\varepsilon^{\mathfrak{ck}}\gen{C}^{\mathfrak{f}}{}_{\mathfrak{k}}.
\]

For $\alg{h}_+$ there is a non-degenerate invariant supersymmetric bilinear form,
so we can write down the quadratic Casimir invariant
\<
\mathcal{T}_{\alg{h}_+}
\eq
\half\gen{R}^c{}_d\gen{R}^d{}_c
-\half\gen{L}^\gamma{}_\delta\gen{L}^\delta{}_\gamma
-\half\gen{B}^{\mathfrak{c}}{}_{\mathfrak{d}}\gen{C}^{\mathfrak{d}}{}_{\mathfrak{c}}
-\half\gen{C}^{\mathfrak{d}}{}_{\mathfrak{c}}\gen{B}^{\mathfrak{c}}{}_{\mathfrak{d}}
-\half\varepsilon_{ad}\varepsilon_{\beta\epsilon}\varepsilon_{\mathfrak{cf}}
\gen{Q}^{a\beta\mathfrak{c}}\gen{Q}^{d\epsilon\mathfrak{f}}
\nln\eq
\mathcal{T}_{\psu}
-\half\gen{B}^{\mathfrak{c}}{}_{\mathfrak{d}}\gen{C}^{\mathfrak{d}}{}_{\mathfrak{c}}
-\half\gen{C}^{\mathfrak{c}}{}_{\mathfrak{d}}\gen{B}^{\mathfrak{d}}{}_{\mathfrak{c}},
\>
where $\mathcal{T}_{\psu}$ is the $\psu$ Casimir defined in \eqref{eq:psucasimir}.
One might also add a term proportional to
$\vec{\gen{C}}^2=\half\gen{C}^{\mathfrak{c}}{}_{\mathfrak{d}}\gen{C}^{\mathfrak{d}}{}_{\mathfrak{c}}$
which is obviously central.
The loop algebra of $\alg{h}_+$ therefore
has the following standard classical r-matrix
\[
r\indup{\alg{h}_+}
=
r\indup{\alg{psu}(2|2)}
-
\sum_{m=0}^\infty
(\gen{B}_{-1-m})^{\mathfrak{c}}{}_{\mathfrak{d}}\otimes(\gen{C}_{m})^{\mathfrak{d}}{}_{\mathfrak{c}}
-
\sum_{m=0}^\infty
(\gen{C}_{-1-m})^{\mathfrak{c}}{}_{\mathfrak{d}}\otimes(\gen{B}_{m})^{\mathfrak{d}}{}_{\mathfrak{c}},
\]
following the construction outlined in \secref{sec:classreview}. 
Henceforth, the r-matrix satisfies the CYBE 
and the corresponding Lie bialgebra is quasi-triangular.

%%%%%%%%%%%%%%%%%%%%%%%%%%%%%%%%%%%%%%%%
\paragraph{Reduction of the Algebra.}

To make contact to physics
we want to work on the fundamental representation of $\alg{h}$.
It is easily seen that the automorphisms cannot
be realised on the fundamental representation,
hence $r_{\alg{h}_+}$ cannot produce the
desired fundamental r-matrix.
Nevertheless there is a reduction of the algebra which
leads to the desired r-matrix.
Mathematically, the reduction consists in two steps:
First we restrict the automorphisms
to a particular subalgebra.
The corresponding subalgebra of $\alg{h}_+$
has an ideal consisting
of linear combinations of the charges
which we then project out.
The resulting algebra is the one introduced in
\secref{sec:algdef}.

In particular, we restrict the automorphisms
$(\gen{B}_n)^{\mathfrak{a}}{}_{\mathfrak{b}}$
to a subalgebra spanned by the linear combinations
\[
\gen{B}_n
=
(\gen{B}_{n})^{1}{}_{1}
-(\gen{B}_{n})^{2}{}_{2}
+
2\alpha^{-1}\beta(\gen{B}_{n-1})^{1}{}_{2}
+
2\alpha\beta(\gen{B}_{n-1})^{2}{}_{1}.
\]
The brackets of these automorphisms
with the supercharges agree with those
derived in \eqref{eq:BQS}.
Furthermore, the brackets with the
charges $(\gen{C}_n)^{\mathfrak{a}}{}_{\mathfrak{b}}$
read
\<
\comm{\gen{B}_m}{(\gen{C}_{n})^{1}{}_{1}-(\gen{C}_{n})^{2}{}_{2}}
\eq
-4\alpha^{-1}\beta(\gen{C}_{m+n-1})^{1}{}_{2}
+4\alpha\beta(\gen{C}_{m+n-1})^{2}{}_{1},
\nln
\comm{\gen{B}_m}{(\gen{C}_{n})^{1}{}_{2}}
\eq
+2(\gen{C}_{m+n})^{1}{}_{2}
-2\alpha\beta(\gen{C}_{m+n-1})^{1}{}_{1}
+2\alpha\beta(\gen{C}_{m+n-1})^{2}{}_{2}
,
\\\nn
\comm{\gen{B}_m}{(\gen{C}_{n})^{2}{}_{1}}
\eq
-2(\gen{C}_{n+m})^{2}{}_{1}
+2\alpha^{-1}\beta(\gen{C}_{m+n-1})^{1}{}_{1}
-2\alpha^{-1}\beta(\gen{C}_{m+n-1})^{2}{}_{2}.
\>

It can be seen that the following linear combinations
of the charges span an ideal of the subalgebra
\[
\beta(\gen{C}_{n})^{1}{}_{1}-\beta(\gen{C}_{n})^{2}{}_{2}-\alpha^{-1}(\gen{C}_{n+1})^{1}{}_{2},
\qquad
\beta(\gen{C}_{n})^{1}{}_{1}-\beta(\gen{C}_{n})^{2}{}_{2}-\alpha(\gen{C}_{n+1})^{2}{}_{1}.
\]
We project out this ideal
analogously to the projection which
turns $\alg{u}(n|n)$ into $\alg{pu}(n|n)$
or $\alg{su}(n|n)$ into $\alg{psu}(n|n)$.
The remaining charges will be denoted by $\gen{C}_n$
defined through
\[\label{eq:Cproject}
(\gen{C}_{n})^{1}{}_{1}=-(\gen{C}_{n})^{2}{}_{2}=\gen{C}_{n},
\qquad
(\gen{C}_{n})^{1}{}_{2}=2\alpha\beta \gen{C}_{n-1},
\qquad
(\gen{C}_{n})^{2}{}_{1}=2\alpha^{-1}\beta\gen{C}_{n-1}.
\]
This step makes the brackets of supercharges
coincide with \eqref{eq:QSbi}
and furthermore the charges $\gen{C}_n$ become central
\[
\comm{\gen{B}_n}{\gen{C}_m}=0.
\]
In conclusion, the reduction of the algebra leads to
the algebra discussed in \secref{sec:algdef}.

%%%%%%%%%%%%%%%%%%%%%%%%%%%%%%%%%%%%%%%%
\paragraph{Reduction of the Coalgebra.}

One can convince oneself that the standard r-matrix
$r\indup{\alg{h}_+}$ for $\alg{h}_+$
contains terms which are not part of the reduced algebra.
We solve this problem by modifying the r-matrix slightly
before preforming the reduction of the algebra.
The modified r-matrix will not be meaningful in the original
algebra, but it will complete the reduced algebra to a
quasi-triangular bialgebra.

Before we perform the reduction of the algebra we
twist the r-matrix according to \eqref{eq:gentwist}
with $\gen{J}:=(\gen{C}_{-1})^{1}{}_{1}$
and $\gen{J}':=(\gen{B}_0)^{1}{}_{1}-(\gen{B}_0)^{2}{}_{2}$
\[
r:=r_{\alg{h}_+}+\gen{J}\wedge\gen{J}'
=r\indup{\alg{h}_+}
+(\gen{C}_{-1})^1{}_1\wedge\bigbrk{(\gen{B}_{0})^{1}{}_{1}-(\gen{B}_{0})^{2}{}_{2}}.
\]
It is clear that $\comm{\gen{J}}{\gen{J}'}=0$,
and because $\gen{J}'$ is a Cartan element we also have
$\comm{\gen{J}'}{r_{\alg{h}_+}}=0$.
However it turns out that
\[\label{eq:cybeproblem}
\comm{\gen{J}}{r\indup{\alg{h_+}}}=
(\gen{C}_{-1})^{1}{}_{2}\wedge (\gen{C}_{-1})^{2}{}_{1},
\]
and therefore the twisted r-matrix $r$ does not satisfy
the CYBE according to \eqref{eq:gentwistcybe}
\[
\cybe{r}{r}=-\gen{J}'\mathbin{\wedge\wedge}\comm{\gen{J}}{r\indup{\alg{h_+}}}
=
-\bigbrk{(\gen{B}_0)^{1}{}_{1}-(\gen{B}_0)^{2}{}_{2}}\wedge(\gen{C}_{-1})^{1}{}_{2}\wedge(\gen{C}_{-1})^{2}{}_{1}
\neq 0.
\]

Now let us consider the reduced algebra.
The projection of the central charges \eqref{eq:Cproject}
has two interesting consequences:
Firstly, it makes
the combination in \eqref{eq:cybeproblem} vanish,
$\comm{\gen{J}}{r\indup{\alg{h_+}}}=0$
and thus it reinstates the CYBE
$\cybe{r}{r}=0$.
Secondly, the twisted r-matrix can be written as
\[
r=
r\indup{\alg{psu}(2|2)}
-\sum_{m=-1}^\infty\gen{B}_{-1-m}\otimes\gen{C}_{m}
-\sum_{m=+1}^\infty\gen{C}_{-1-m}\otimes\gen{B}_{m},
\]
i.e.\ all the undesired combinations of $(\gen{B}_n)^{\mathfrak{a}}{}_{\mathfrak{b}}$
which are not part of the reduced algebra have dropped out.
This classical r-matrix fully agrees with
our proposal \eqref{eq:rclass} including
the shifted bounds of the sums.
Thus the reduced bialgebra is identical to the one considered in
\secref{sec:algdef}.

A similar construction may be possible
for the exceptional loop algebra of $\alg{d}(2,1;\varepsilon)$
in the limit $\varepsilon\to0$.
It is worth pursuing this question further.

%%%%%%%%%%%%%%%%%%%%%%%%%%%%%%%%%%%%%%%%%%%%%%%%%%%%%%%%%%%%%%%%%%%%%%%%%%%%%%%%
%%%%%%%%%%%%%%%%%%%%%%%%%%%%%%%%%%%%%%%%%%%%%%%%%%%%%%%%%%%%%%%%%%%%%%%%%%%%%%%%
\section{Different Classical Limits}\label{sec:weak}

In this section we investigate the behaviour
of the R-matrix in different limits.
Recall that for ordinary Yangians the R-matrix depends only on one variable,
the difference of the spectral parameters.
Requiring that $\rmat \to 1\otimes 1 $ basically defines this limit uniquely:
the difference of spectral parameters must approach infinity.
To investigate the classical limit one introduces
an unphysical scaling parameter $\hbar$
such that the spectral parameters scale like $u\sim \hbar^{-1}$.
In contradistinction, our R-matrix does not
only depend on the spectral parameters, but also
on the physical coupling constant $g$.
In the previous sections we used $\hbar = g^{-1}$ as a natural scale,
and let the spectral parameter scale like $u\sim \hbar^{-1}$.
However, there are other consistent ways of rescaling $u$ and $g$
by functions of $\hbar$,
which makes it possible to have several well-defined classical limits.
This is reminiscent of the AdS/CFT correspondence,
which has a classical limit at strong and at weak coupling.
It is even possible to define different classical limits within
the strong or weak coupling regime, respectively.
Our algebraic framework might make
it possible to find other interesting classical limits or even to classify them.

For a different classical limit
we define the natural scale by $g = \hbar^{-\kappa}$ with $\kappa<1$,
and let the spectral parameters scale as usual, $u\sim \hbar^{-1}$,
and also $x\sim \hbar^{-1}$.
Similarly, we could let $x\sim \hbar$
which would lead to qualitatively the same results.
Thus we are in the weak coupling regime for $\kappa<0$
and in the strong coupling regime for $0<\kappa<1$;
nevertheless the limit will not make a distinction
between positive and negative $\kappa$.
We introduce the rescaled spectral parameter
\[
 \tu{} = u \hbar.
\]
We also choose as the phase factor the same as in \eqref{eq:phaseLC},
leading to a classical r-matrix taking the same
form as given at the top of \tabref{tab:rCoeff},
but with the coefficients taking the values
\<
\half A_{12}+\half B_{12}=\half D_{12}+\half E_{12}
=\frac{\gat{2}}{\gat{1}}H_{12}=\frac{\gat{1}}{\gat{2}}K_{12}\eq\frac{1}{i\tilde{u}_1-i\tu{2}}\,,\nln
\half A_{12}-\half B_{12}=\half D_{12}-\half E_{12}\eq -\frac{1}{4i\tu{1}}+\frac{1}{4i\tu{2}}\,,\nln
C_{12}=F_{12}\eq 0,\nln
G_{12}=-L_{12}\eq +\frac{1}{4i\tu{1}}+\frac{1}{4i\tu{2}}\,.
\>
If we use the phase factor \eqref{eq:phaseexactmom} instead of \eqref{eq:phaseLC}
we have to add to $r$ the diagonal terms
$(1\otimes 1)r_0$ with
\[
r_0\simeq \frac{1}{4i\tu{2}}-\frac{1}{4i\tu{1}}\,.
\]

The above r-matrix can be written compactly
as
\[\label{eq:stdfundr}
r\simeq \frac{\perm_{12}}{i\tu{1}-i\tu{2}}+
\gen{C}_{-1}\wedge\gen{B}_{0},
\]
where $\perm_{12}$ is the (graded) permutation.
This action coincides with the action of the
same r-matrix \eqref{eq:rclass} discussed in \secref{sec:u22def}
if one sets the parameters for the fundamental representation in
\eqref{eq:fermigen} to
\[
 a = \frac{1}{d} = \gat{}, \qquad b = c = 0 .
\]
This fundamental representation is obviously consistent with the
undeformed loop algebra $\alg{u}(2|2)[i\tu{},(i\tu{})^{-1}]$
(i.e.\ the deformed algebra with $\beta=0$).%
\footnote{In other words, the $\beta=0$ undeformed algebra is a contraction
of the $\beta=1$ deformed algebra where one scales the level-$n$ generators
$\gen{J}_n$ by $\epsilon^n$ and takes the $\epsilon\to0$ limit.}
Effectively this means that we reproduce the standard fundamental r-matrix
for $\alg{u}(2|2)[i\tu{},(i\tu{})^{-1}]$ with
a Reshetikhin twist \eqref{eq:gentwist}
which takes the form \eqref{eq:stdfundr}.

%%%%%%%%%%%%%%%%%%%%%%%%%%%%%%%%%%%%%%%%%%%%%%%%%%%%%%%%%%%%%%%%%%%%%%%%%%%%%%%%
%%%%%%%%%%%%%%%%%%%%%%%%%%%%%%%%%%%%%%%%%%%%%%%%%%%%%%%%%%%%%%%%%%%%%%%%%%%%%%%%
\section{Lift to Hopf Algebra}
\label{sec:lift}

Let us briefly discuss the lift of the Lie bialgebra to a Hopf algebra.
We shall only consider the fundamental evaluation representation, 
without proving that the relations we give lead to a consistent Hopf algebra.
In this case the generators act like $\gen{J}^A_n\simeq (iu)^n\gen{J}^A_0$, 
with $\gen{J}^A_0$ in the fundamental representation of the Lie algebra.
The action of most of the generators is known from
\eqref{eq:bosegen,eq:fermigen}.
In order to determine the action of $\gen{B}$ and its
commutation relations, we have made an ansatz similar to
\eqref{eq:Bclassfund} and \eqref{eq:BQS} with undetermined
coefficients.
It turns out that
\[
\gen{B}\state{\phi^a}=-\frac{1}{4C}\,\state{\phi^a},
\qquad
\gen{B}\state{\psi^\alpha}=+\frac{1}{4C}\,\state{\psi^\alpha}.
\]
is compatible with the commutators
\<\label{eq:BQShopf}
\bigcomm{\gen{B}_m}{(\gen{Q}_n){}^\alpha{}_b}\earel{\simeq}
(\gen{Q}_{m+n}){}^\alpha{}_b
-\alpha\varepsilon^{\alpha\gamma}\varepsilon_{bd}(1+\eip^2)(\gen{S}_{m+n-1}){}^d{}_\gamma,
\nln
\bigcomm{\gen{B}_m}{(\gen{S}_n){}^a{}_\beta}\earel{\simeq}
-(\gen{S}_{m+n}){}^a{}_\beta
-\alpha^{-1}\varepsilon^{ac}\varepsilon_{\beta\delta}(1+\eip^{-2})(\gen{Q}_{m+n-1}){}^\delta{}_c.
\>
Furthermore we can write the two additional
central charges $\gen{P},\gen{K}$ appearing in the
commutators of alike supercharges
using the charge $\gen{C}$ at a lower level
\<
\bigacomm{(\gen{Q}_m){}^{\alpha}{}_{b}}{(\gen{Q}_n){}^{\gamma}{}_{d}}\earel{\simeq}
\alpha\varepsilon^{\alpha\gamma}\varepsilon_{bd}(1+\eip^2)\gen{C}_{m+n-1},
\nln
\bigacomm{(\gen{S}_m){}^{a}{}_{\beta}}{(\gen{S}_n){}^{c}{}_{\delta}}\earel{\simeq}
-\alpha^{-1}\varepsilon^{ac}\varepsilon_{\beta\delta}(1+\eip^{-2})\gen{C}_{m+n-1}.
\>
This would lead to the following relations between $\gen{C}_{-1}$ and $\eip^2$
\[\gen{C}_{-1}\simeq g\,\frac{1-\eip^2}{1+\eip^2}\,,\qquad
\eip^2\simeq \frac{g-\gen{C}_{-1}}{g+\gen{C}_{-1}}\,.\]
Note that the above relations hold only on the fundamental evaluation
representation. There may be further corrections which cannot be
seen on this representation.

Next we would like to see if the generators
$\gen{B}_n$ are symmetries of the R-matrix.
Due to the relation \eqref{eq:doubleyang}
for evaluation representations
it suffices to consider $\gen{B}_0$ and $\gen{B}_1$.
The cobrackets in \tabref{tab:cobra} of these generators read%
\footnote{Note that $\sum_{k=1}^{-1}f_k=-f_0$.}
\<
\coprocl(\gen{B}_0)\eq
-\alpha^{-1}\varepsilon^{bd}\varepsilon_{\alpha\gamma}
(\gen{Q}_{-1})^\alpha{}_b\wedge(\gen{Q}_{-1})^\gamma{}_d
+\alpha\varepsilon^{\beta\delta}\varepsilon_{ac}
(\gen{S}_{-1})^a{}_\beta\wedge(\gen{S}_{-1})^c{}_\delta,
\nln
\coprocl(\gen{B}_1)\eq
(\gen{Q}_0)^\alpha{}_b\wedge(\gen{S}_0)^b{}_\alpha.
\>

There is a natural lift of the cobracket for $\gen{B}_1$ to
a coproduct. One merely has to add the standard coproduct
and introduce proper braiding factors for $\gen{Q}$ and $\gen{S}$
\[
\copro(\gen{B}_1)=
\gen{B}_1\otimes 1+1\otimes\gen{B}_1
+\half
g^{-1}(\gen{Q}_0)^\alpha{}_b\eip^{-1}\otimes(\gen{S}_0)^b{}_\alpha
+\half
g^{-1}(\gen{S}_0)^a{}_\beta\eip^{+1}\otimes(\gen{Q}_0)^\beta{}_a.
\]
This coproduct was proposed very recently
and independently in \cite{Matsumoto:2007rh}
where it was also shown to be a symmetry of the fundamental S-matrix.
Here, we confirm this result which gives
further confidence that our Lie bialgebra has
a lift to a quasi-triangular double Yangian
with the known fundamental R-matrix.
We furthermore expect that the four new fermionic coproducts proposed
in the conclusions of \cite{Matsumoto:2007rh}
are linear combinations of the coproducts
of our generators $\gen{Q}_{0,1}$ and $\gen{S}_{0,1}$.
The reason for the discrepancy lies in different assumed algebra relations,
\eqref{eq:BQShopf} in our case and
those related to the representation discussed in \secref{sec:MT}
for \cite{Matsumoto:2007rh}.

Finding a coproduct for $\gen{B}_0$ is however not as easy:
According to \eqref{eq:grading}
the grading of the two terms in $\delta(\gen{B}_0)$
is $\pm 2$ whereas the grading of $\gen{B}_0$ should be zero.
This mismatch leads to inconsistencies in the braiding with $\eip$
and to a failure of coassociativity in the coproduct.
It is currently not clear how to resolve this issue.
To have a coproduct for all $\gen{B}_n$ is nevertheless important
for the lift of the classical r-matrix to a universal R-matrix.

%%%%%%%%%%%%%%%%%%%%%%%%%%%%%%%%%%%%%%%%%%%%%%%%%%%%%%%%%%%%%%%%%%%%%%%%%%%%%%%%
%%%%%%%%%%%%%%%%%%%%%%%%%%%%%%%%%%%%%%%%%%%%%%%%%%%%%%%%%%%%%%%%%%%%%%%%%%%%%%%%
\section{Conclusions and Outlook}

In this paper we have proposed a quasi-triangular
Lie bialgebra whose underlying Lie algebra 
is a deformation of the loop algebra of $\alg{u}(2|2)$.
Its classical r-matrix on the fundamental evaluation representation
coincides with the classical limit of the S-matrix of \cite{Beisert:2006zy}
obtained in \cite{Torrielli:2007mc}.
This bialgebra is almost the standard rational loop bialgebra
based on $\alg{u}(2|2)$, but there are two crucial differences:
Firstly, the cobrackets include some additional non-standard terms,
even for some level-0 generators.
Secondly, not all Lie brackets have a uniform level, there is mixing
between the levels.
In comparison to the Hopf algebra symmetries
of the S-matrix \cite{Gomez:2006va,Plefka:2006ze,Beisert:2007ds}
which use the three-dimensional universal central extension of $\psu$,
here we have only one central extension $\gen{C}$.
The three central elements of $\psucentral$ clearly have no dual
partners (automorphisms) which would be needed
for the standard construction of the classical r-matrix.
Conversely, our central element $\gen{C}$ can be paired with
the $\alg{u}(1)$ automorphism $\gen{B}$ of $\alg{u}(2|2)$
as in the proposal of \cite{Moriyama:2007jt},
and we can construct a classical r-matrix.
The additional central charges appear in our algebra 
as the central charge at a different loop level.
Furthermore, the $\alg{u}(1)$ automorphism 
does not act diagonally on the roots,
and mixes the levels of the loop algebra.
Due to these features the Lie algebra does not
coincide with the standard $\alg{u}(2|2)$ loop algebra, 
but to some extent the two can be related, see \secref{sec:algu22}.
Apart from that and unlike in \cite{Moriyama:2007jt},
our bialgebra resembles the standard one for Yangian doubles.
Since \cite{Moriyama:2007jt} reproduces the correct classical r-matrix
on the fundamental representation,
it would be interesting to find out if their proposal and ours are equivalent.
For instance, one might compare the two when acting on bound states
\cite{Dorey:2006dq,Chen:2006gq,Roiban:2006gs,Chen:2006gp,Beisert:2006qh}.
Whether or not they are equivalent, we believe that our approach is more suitable
to find the quantisation of the bialgebra to a double Yangian Hopf algebra.
The latter should be equipped with a universal R-matrix,
which should be almost of the standard form for $\grp{DY}(\alg{u}(2|2))$.
Again, the crucial difference will be the behaviour of the $\alg{u}(1)$ automorphism
and the level-mixing due to the identifications between the central charges,
the braiding and the loop variable.
We have made first steps in this direction in the previous section,
where we have found a hidden symmetry of the
fundamental R-matrix which was independently discovered
in the recent paper \cite{Matsumoto:2007rh}
(which appeared while we were preparing our manuscript).
However there are some unresolved issues concerning the
full quantum braiding for the coproduct of one remaining generator.
It would furthermore be interesting to see if contact with
quantum deformations and the exceptional Lie superalgebra $\alg{d}(2,1;\alpha)$
\cite{Zhang:1992aa,Yamane:1994aa,Yamane:1999aa,Yamane:2003aa,Heckenberger:2007aa,Beisert:2008tw},
or, on a different account, with the loop algebras
encountered in the context of a twistor formulation
of $\superN=4$ SYM \cite{Wolf:2004hp,Popov:2006qu},
can be made.
It would also be desirable to understand the bialgebra structure
for other kinematical regimes such as giant magnons
\cite{Hofman:2006xt} and the near-flat limit
\cite{Maldacena:2006rv,Klose:2006zd,Klose:2007rz}
and how they are related to our near plane wave setup.
Concerning the algebraic determination of the dressing phase,
we find no constraint for the classical r-matrix.
However, quasi-triangularity for the universal R-matrix
leads to stronger constraints and may allow a derivation
from first principles.

%%%%%%%%%%%%%%%%%%%%%%%%%%%%%%%%%%%%%%%%
\paragraph{Acknowledgements.}

We would like to thank
Valentina Forini, Peter Koroteev, Tristan McLoughlin, Jan Plefka,
Alessandro Torrielli
and the CMP referee
for interesting discussions and useful comments.

%%%%%%%%%%%%%%%%%%%%%%%%%%%%%%%%%%%%%%%%
\bibliography{classr}

\begin{thebibliography}{10}
\ifx\href\asklfhas\newcommand{\href}[2]{#2}\fi
\ifx\arxivref\asklfhas\newcommand{\arxivref}[2]{\href{http://arxiv.org/abs/#1}%
{#2}}\fi
\ifx\doiref\asklfhas\newcommand{\doiref}[2]{\href{http://dx.doi.org/#1}{#2}}\fi
\raggedright
\small
\parskip 0pt

%%CITATION = HEP-TH 0212208;%%
\bibitem{Minahan:2002ve}
J.~A.~Minahan and K.~Zarembo,
\textit{``The Bethe-ansatz for {$\mathcal{N}=\mathord{}$4} super Yang-Mills''},
\textsf{\doiref{10.1088/1126-6708/2003/03/013}{JHEP~0303,~013~(2003)}},
\texttt{\arxivref{hep-th/0212208}{hep-th/0212208}}.
%
%%CITATION = HEP-TH 0303060;%%
\bibitem{Beisert:2003tq}
N.~Beisert, C.~Kristjansen and M.~Staudacher,
\textit{``The Dilatation Operator of {$\mathcal{N}=\mathord{}$4} Conformal
  Super Yang-Mills Theory''},
\textsf{\doiref{10.1016/S0550-3213(03)00406-1}{Nucl.~Phys.~B664,~131~(2003)}},
\texttt{\arxivref{hep-th/0303060}{hep-th/0303060}}.
%
%%CITATION = HEP-TH 0305116;%%
\bibitem{Bena:2003wd}
I.~Bena, J.~Polchinski and R.~Roiban,
\textit{``Hidden symmetries of the {$AdS_5\times S^5$} superstring''},
\textsf{\doiref{10.1103/PhysRevD.69.046002}{Phys.~Rev.~D69,~046002~(2004)}},
\texttt{\arxivref{hep-th/0305116}{hep-th/0305116}}.
%
%%CITATION = HEP-TH 0307042;%%
\bibitem{Beisert:2003yb}
N.~Beisert and M.~Staudacher,
\textit{``The {$\mathcal{N}=\mathord{}$4} SYM Integrable Super Spin Chain''},
\textsf{\doiref{10.1016/j.nuclphysb.2003.08.015}{Nucl.~Phys.~B670,~439~(2003)}%
},
\texttt{\arxivref{hep-th/0307042}{hep-th/0307042}}.
%
%%CITATION = HEP-TH 0504190;%%
\bibitem{Beisert:2005fw}
N.~Beisert and M.~Staudacher,
\textit{``Long-Range PSU(2,2$/$4) Bethe Ansaetze for Gauge Theory and
  Strings''},
\textsf{\doiref{10.1016/j.nuclphysb.2005.06.038}{Nucl.~Phys.~B727,~1~(2005)}},
\texttt{\arxivref{hep-th/0504190}{hep-th/0504190}}.
%
%%CITATION = HEP-TH 0412188;%%
\bibitem{Staudacher:2004tk}
M.~Staudacher,
\textit{``The factorized S-matrix of CFT/AdS''},
\textsf{\doiref{10.1088/1126-6708/2005/05/054}{JHEP~0505,~054~(2005)}},
\texttt{\arxivref{hep-th/0412188}{hep-th/0412188}}.
%
%%CITATION = HEP-TH/0511082;%%
\bibitem{Beisert:2005tm}
N.~Beisert,
\textit{``The su(2$/$2) dynamic S-matrix''},
\texttt{\arxivref{hep-th/0511082}{hep-th/0511082}}.
%
%%CITATION = HEP-TH 0609157;%%
\bibitem{Arutyunov:2006ak}
G.~Arutyunov, S.~Frolov, J.~Plefka and M.~Zamaklar,
\textit{``The Off-shell Symmetry Algebra of the Light-cone $AdS_5\times S^5$
  Superstring''},
\textsf{\doiref{10.1088/1751-8113/40/13/018}{J.~Phys.~A40,~3583~(2007)}},
\texttt{\arxivref{hep-th/0609157}{hep-th/0609157}}.
%
%%CITATION = HEP-TH 0608029;%%
\bibitem{Gomez:2006va}
C.~Gomez and R.~Hern\'andez,
\textit{``The magnon kinematics of the AdS/CFT correspondence''},
\textsf{\doiref{10.1088/1126-6708/2006/11/021}{JHEP~0611,~021~(2006)}},
\texttt{\arxivref{hep-th/0608029}{hep-th/0608029}}.
%
%%CITATION = HEP-TH 0608038;%%
\bibitem{Plefka:2006ze}
J.~Plefka, F.~Spill and A.~Torrielli,
\textit{``On the Hopf algebra structure of the AdS/CFT S-matrix''},
\textsf{\doiref{10.1103/PhysRevD.74.066008}{Phys.~Rev.~D74,~066008~(2006)}},
\texttt{\arxivref{hep-th/0608038}{hep-th/0608038}}.
%
%%CITATION = NLIN.SI 0610017;%%
\bibitem{Beisert:2006qh}
N.~Beisert,
\textit{``The Analytic Bethe Ansatz for a Chain with Centrally Extended
  su(2$/$2) Symmetry''},
\textsf{\doiref{10.1088/1742-5468/2007/01/P01017}{J.~Stat.~Mech.~07,~P01017~(2%
007)}},
\texttt{\arxivref{nlin.SI/0610017}{nlin.SI/0610017}}.
%
%%CITATION = HEP-TH 0401243;%%
\bibitem{Dolan:2004ps}
L.~Dolan, C.~R.~Nappi and E.~Witten,
\textit{``Yangian symmetry in $D=$4 superconformal Yang-Mills theory''},
\texttt{\arxivref{hep-th/0401243}{hep-th/0401243}},
in: \textit{``Quantum Theory and Symmetries''},
ed.: P.~C.~Argyres et~al.,
World Scientific (2004),
Singapore.
%
%%CITATION = HEP-TH 0308089;%%
\bibitem{Dolan:2003uh}
L.~Dolan, C.~R.~Nappi and E.~Witten,
\textit{``A Relation Between Approaches to Integrability in Superconformal
  Yang-Mills Theory''},
\textsf{\doiref{10.1088/1126-6708/2003/10/017}{JHEP~0310,~017~(2003)}},
\texttt{\arxivref{hep-th/0308089}{hep-th/0308089}}.
%
%%CITATION = HEP-TH 0401057;%%
\bibitem{Serban:2004jf}
D.~Serban and M.~Staudacher,
\textit{``Planar {$\mathcal{N}=\mathord{}$4} gauge theory and the Inozemtsev
  long range spin chain''},
\textsf{\doiref{10.1088/1126-6708/2004/06/001}{JHEP~0406,~001~(2004)}},
\texttt{\arxivref{hep-th/0401057}{hep-th/0401057}}.
%
%%CITATION = HEP-TH 0409180;%%
\bibitem{Agarwal:2004sz}
A.~Agarwal and S.~G.~Rajeev,
\textit{``Yangian symmetries of matrix models and spin chains: The dilatation
  operator of {$\mathcal{N}=\mathord{}$4} SYM''},
\textsf{\doiref{10.1142/S0217751X05022822}{Int.~J.~Mod.~Phys.~A20,~5453~(2005)%
}},
\texttt{\arxivref{hep-th/0409180}{hep-th/0409180}}.
%
%%CITATION = HEP-TH 0610283;%%
\bibitem{Zwiebel:2006cb}
B.~I.~Zwiebel,
\textit{``Yangian symmetry at two-loops for the su(2$/$1) sector of
  {$\mathcal{N}=\mathord{}$4} SYM''},
\textsf{\doiref{10.1088/1751-8113/40/5/018}{J.~Phys.~A40,~1141~(2007)}},
\texttt{\arxivref{hep-th/0610283}{hep-th/0610283}}.
%
%%CITATION = ARXIV:0704.0400;%%
\bibitem{Beisert:2007ds}
N.~Beisert,
\textit{``The S-Matrix of AdS/CFT and Yangian Symmetry''},
\textsf{PoS~Solvay,~002~(2007)},
\texttt{\arxivref{0704.0400}{arxiv:0704.0400}}.
%
%%CITATION = HEP-TH/0612229;%%
\bibitem{Arutyunov:2006yd}
G.~Arutyunov, S.~Frolov and M.~Zamaklar,
\textit{``The Zamolodchikov-Faddeev algebra for $AdS_5\times S^5$
  superstring''},
\textsf{\doiref{10.1088/1126-6708/2007/04/002}{JHEP~0704,~002~(2007)}},
\texttt{\arxivref{hep-th/0612229}{hep-th/0612229}}.
%
%%CITATION = HEP-TH/0701200;%%
\bibitem{Gomez:2007zr}
C.~Gomez and R.~Hern\'andez,
\textit{``Quantum deformed magnon kinematics''},
\textsf{\doiref{10.1088/1126-6708/2007/03/108}{JHEP~0703,~108~(2007)}},
\texttt{\arxivref{hep-th/0701200}{hep-th/0701200}}.
%
%%CITATION = ARXIV:0704.2069;%%
\bibitem{Young:2007wd}
C.~A.~S.~Young,
\textit{``q-Deformed Supersymmetry and Dynamic Magnon Representations''},
\textsf{\doiref{10.1088/1751-8113/40/30/033}{J.~Phys.~A40,~9165~(2007)}},
\texttt{\arxivref{0704.2069}{arxiv:0704.2069}}.
%
%%CITATION = ARXIV:0707.1031;%%
\bibitem{Beisert:2007sk}
N.~Beisert and B.~I.~Zwiebel,
\textit{``On Symmetry Enhancement in the psu(1,1$/$2) Sector of
  {$\mathcal{N}=\mathord{}$4} SYM''},
\textsf{\doiref{10.1088/1126-6708/2007/10/031}{JHEP~0710,~031~(2007)}},
\texttt{\arxivref{0707.1031}{arxiv:0707.1031}}.
%
%%CITATION = HEP-TH 0603038;%%
\bibitem{Janik:2006dc}
R.~A.~Janik,
\textit{``The $AdS_5\times S^5$ superstring worldsheet S-matrix and crossing
  symmetry''},
\textsf{\doiref{10.1103/PhysRevD.73.086006}{Phys.~Rev.~D73,~086006~(2006)}},
\texttt{\arxivref{hep-th/0603038}{hep-th/0603038}}.
%
%%CITATION = HEP-TH 0609044;%%
\bibitem{Beisert:2006ib}
N.~Beisert, R.~Hern\'andez and E.~L\'opez,
\textit{``A Crossing-Symmetric Phase for $AdS_5 \times S^5$ Strings''},
\textsf{\doiref{10.1088/1126-6708/2006/11/070}{JHEP~0611,~070~(2006)}},
\texttt{\arxivref{hep-th/0609044}{hep-th/0609044}}.
%
%%CITATION = HEP-TH 0610251;%%
\bibitem{Beisert:2006ez}
N.~Beisert, B.~Eden and M.~Staudacher,
\textit{``Transcendentality and crossing''},
\textsf{\doiref{10.1088/1742-5468/2007/01/P01021}{J.~Stat.~Mech.~07,~P01021~(2%
007)}},
\texttt{\arxivref{hep-th/0610251}{hep-th/0610251}}.
%
%%CITATION = HEP-TH/9406194;%%
\bibitem{Khoroshkin:1994uk}
S.~M.~Khoroshkin and V.~N.~Tolstoi,
\textit{``Yangian double''},
\textsf{\doiref{10.1007/BF00714404}{Lett.~Math.~Phys.~36,~373~(1996)}},
\texttt{\arxivref{hep-th/9406194}{hep-th/9406194}}.
%
%%CITATION = HEP-TH/0701281;%%
\bibitem{Torrielli:2007mc}
A.~Torrielli,
\textit{``Classical r-matrix of the su(2$/$2) SYM spin-chain''},
\textsf{\doiref{10.1103/PhysRevD.75.105020}{Phys.~Rev.~D75,~105020~(2007)}},
\texttt{\arxivref{hep-th/0701281}{hep-th/0701281}}.
%
%%CITATION = HEP-TH/0611169;%%
\bibitem{Klose:2006zd}
T.~Klose, T.~McLoughlin, R.~Roiban and K.~Zarembo,
\textit{``Worldsheet scattering in $AdS_5\times S^5$''},
\textsf{\doiref{10.1088/1126-6708/2007/03/094}{JHEP~0703,~094~(2007)}},
\texttt{\arxivref{hep-th/0611169}{hep-th/0611169}}.
%
%%CITATION = ARXIV:0704.3891;%%
\bibitem{Klose:2007rz}
T.~Klose, T.~McLoughlin, J.~A.~Minahan and K.~Zarembo,
\textit{``World-sheet scattering in $AdS_5\times S^5$ at two loops''},
\textsf{\doiref{10.1088/1126-6708/2007/08/051}{JHEP~0708,~051~(2007)}},
\texttt{\arxivref{0704.3891}{arxiv:0704.3891}}.
%
%%CITATION = JHEPA,0706,083;%%
\bibitem{Moriyama:2007jt}
S.~Moriyama and A.~Torrielli,
\textit{``A Yangian Double for the AdS/CFT Classical r-matrix''},
\textsf{\doiref{10.1088/1126-6708/2007/06/083}{JHEP~0706,~083~(2007)}},
\texttt{\arxivref{0706.0884}{arxiv:0706.0884}}.
%
\bibitem{Chari:1994pz}
V.~Chari and A.~Pressley,
\textit{``A guide to quantum groups''},
Cambridge University Press (1994),
Cambridge, UK,
651p.
%
%%CITATION = JMTSE,41,898;%%
\bibitem{Drinfeld:1986in}
V.~G.~Drinfeld,
\textit{``Quantum groups''},
\textsf{\doiref{10.1007/BF01247086}{J.~Math.~Sci.~41,~898~(1988)}}.
%
%%CITATION = HEP-TH 0407277;%%
\bibitem{Beisert:2004ry}
N.~Beisert,
\textit{``The Dilatation Operator of {$\mathcal{N}=\mathord{}$4} Super
  Yang-Mills Theory and Integrability''},
\textsf{\doiref{10.1016/j.physrep.2004.09.007}{Phys.~Rept.~405,~1~(2004)}},
\texttt{\arxivref{hep-th/0407277}{hep-th/0407277}}.
%
%%CITATION = HEP-TH/0703086;%%
\bibitem{Martins:2007hb}
M.~J.~Martins and C.~S.~Melo,
\textit{``The Bethe ansatz approach for factorizable centrally extended
  S-matrices''},
\textsf{\doiref{10.1016/j.nuclphysb.2007.05.021}{Nucl.~Phys.~B785,~246~(2007)}%
},
\texttt{\arxivref{hep-th/0703086}{hep-th/0703086}}.
%
%%CITATION = HEP-TH 0406256;%%
\bibitem{Arutyunov:2004vx}
G.~Arutyunov, S.~Frolov and M.~Staudacher,
\textit{``Bethe ansatz for quantum strings''},
\textsf{\doiref{10.1088/1126-6708/2004/10/016}{JHEP~0410,~016~(2004)}},
\texttt{\arxivref{hep-th/0406256}{hep-th/0406256}}.
%
%%CITATION = HEP-TH 0509084;%%
\bibitem{Beisert:2005cw}
N.~Beisert and A.~A.~Tseytlin,
\textit{``On Quantum Corrections to Spinning Strings and Bethe Equations''},
\textsf{\doiref{10.1016/j.physletb.2005.09.054}{Phys.~Lett.~B629,~102~(2005)}},
\texttt{\arxivref{hep-th/0509084}{hep-th/0509084}}.
%
%%CITATION = HEP-TH 0603204;%%
\bibitem{Hernandez:2006tk}
R.~Hern{\'a}ndez and E.~L{\'o}pez,
\textit{``Quantum corrections to the string Bethe ansatz''},
\textsf{\doiref{10.1088/1126-6708/2006/07/004}{JHEP~0607,~004~(2006)}},
\texttt{\arxivref{hep-th/0603204}{hep-th/0603204}}.
%
%%CITATION = HEP-TH 0703191;%%
\bibitem{Gromov:2007aq}
N.~Gromov and P.~Vieira,
\textit{``The $AdS_5\times S^5$ superstring quantum spectrum from the algebraic
  curve''},
\textsf{\doiref{10.1016/j.nuclphysb.2007.07.032}{Nucl.~Phys.~B789,~175~(2008)}%
},
\texttt{\arxivref{hep-th/0703191}{hep-th/0703191}}.
%
%%CITATION = HEP-TH 0603008;%%
\bibitem{Frolov:2006cc}
S.~Frolov, J.~Plefka and M.~Zamaklar,
\textit{``The $AdS_5\times S^5$ superstring in light-cone gauge and its Bethe
  equations''},
\textsf{\doiref{10.1088/0305-4470/39/41/S15}{J.~Phys.~A39,~13037~(2006)}},
\texttt{\arxivref{hep-th/0603008}{hep-th/0603008}}.
%
%%CITATION = HEP-TH 0202021;%%
\bibitem{Berenstein:2002jq}
D.~Berenstein, J.~M.~Maldacena and H.~Nastase,
\textit{``Strings in flat space and pp waves from {$\mathcal{N}=\mathord{}$4}
  {Super} {Yang Mills}''},
\textsf{\doiref{10.1088/1126-6708/2002/04/013}{JHEP~0204,~013~(2002)}},
\texttt{\arxivref{hep-th/0202021}{hep-th/0202021}}.
%
%%CITATION = HEP-TH 0208010;%%
\bibitem{Parnachev:2002kk}
A.~Parnachev and A.~V.~Ryzhov,
\textit{``Strings in the near plane wave background and AdS/CFT''},
\textsf{\doiref{10.1088/1126-6708/2002/10/066}{JHEP~0210,~066~(2002)}},
\texttt{\arxivref{hep-th/0208010}{hep-th/0208010}}.
%
%%CITATION = HEP-TH 0307032;%%
\bibitem{Callan:2003xr}
C.~G.~Callan,~Jr., H.~K.~Lee, T.~McLoughlin, J.~H.~Schwarz, I.~Swanson and
  X.~Wu,
\textit{``Quantizing string theory in $AdS_5\times S^5$: Beyond the pp-wave''},
\textsf{\doiref{10.1016/j.nuclphysb.2003.09.008}{Nucl.~Phys.~B673,~3~(2003)}},
\texttt{\arxivref{hep-th/0307032}{hep-th/0307032}}.
%
%%CITATION = HEP-TH 0304255;%%
\bibitem{Frolov:2003qc}
S.~Frolov and A.~A.~Tseytlin,
\textit{``Multi-spin string solutions in {$AdS_5\times S^5$}''},
\textsf{\doiref{10.1016/S0550-3213(03)00580-7}{Nucl.~Phys.~B668,~77~(2003)}},
\texttt{\arxivref{hep-th/0304255}{hep-th/0304255}}.
%
%%CITATION = HEP-TH 0306139;%%
\bibitem{Beisert:2003xu}
N.~Beisert, J.~A.~Minahan, M.~Staudacher and K.~Zarembo,
\textit{``Stringing Spins and Spinning Strings''},
\textsf{\doiref{10.1088/1126-6708/2003/09/010}{JHEP~0309,~010~(2003)}},
\texttt{\arxivref{hep-th/0306139}{hep-th/0306139}}.
%
%%CITATION = HEP-TH 0311203;%%
\bibitem{Kruczenski:2003gt}
M.~Kruczenski,
\textit{``Spin chains and string theory''},
\textsf{\doiref{10.1103/PhysRevLett.93.161602}{Phys.~Rev.~Lett.~93,~161602~(20%
04)}},
\texttt{\arxivref{hep-th/0311203}{hep-th/0311203}}.
%
%%CITATION = HEP-TH 0402207;%%
\bibitem{Kazakov:2004qf}
V.~A.~Kazakov, A.~Marshakov, J.~A.~Minahan and K.~Zarembo,
\textit{``Classical/quantum integrability in AdS/CFT''},
\textsf{\doiref{10.1088/1126-6708/2004/04/024}{JHEP~0405,~024~(2004)}},
\texttt{\arxivref{hep-th/0402207}{hep-th/0402207}}.
%
%%CITATION = HEP-TH 0604043;%%
\bibitem{Arutyunov:2006iu}
G.~Arutyunov and S.~Frolov,
\textit{``On $AdS_5\times S^5$ string S-matrix''},
\textsf{\doiref{10.1016/j.physletb.2006.06.064}{Phys.~Lett.~B639,~378~(2006)}},
\texttt{\arxivref{hep-th/0604043}{hep-th/0604043}}.
%
%%CITATION = HEP-TH 0502226;%%
\bibitem{Beisert:2005bm}
N.~Beisert, V.~Kazakov, K.~Sakai and K.~Zarembo,
\textit{``The Algebraic Curve of Classical Superstrings on $AdS_5\times
  S^5$''},
\textsf{\doiref{10.1007/s00220-006-1529-4}{Commun.~Math.~Phys.~263,~659~(2006)%
}},
\texttt{\arxivref{hep-th/0502226}{hep-th/0502226}}.
%
%%CITATION = LMPHD,20,331;%%
\bibitem{Reshetikhin:1990ep}
N.~Reshetikhin,
\textit{``Multiparameter quantum groups and twisted quasitriangular Hopf
  algebras''},
\textsf{\doiref{10.1007/BF00626530}{Lett.~Math.~Phys.~20,~331~(1990)}}.
%
\bibitem{Serganova:1985aa}
V.~V.~Serganova,
\textit{``Automorphisms of simple Lie superalgebras''},
\textsf{\doiref{10.1070/IM1985v024n03ABEH001250}{Math.~USSR~Izv.~24,~539~(1985%
)}}.
%
%%CITATION = ARXIV:0708.1285;%%
\bibitem{Matsumoto:2007rh}
T.~Matsumoto, S.~Moriyama and A.~Torrielli,
\textit{``{A Secret Symmetry of the AdS/CFT S-matrix}''},
\textsf{\doiref{10.1088/1126-6708/2007/09/099}{JHEP~0709,~099~(2007)}},
\texttt{\arxivref{0708.1285}{arxiv:0708.1285}}.
%
%%CITATION = HEP-TH 0606214;%%
\bibitem{Beisert:2006zy}
N.~Beisert,
\textit{``On the scattering phase for $AdS_5\times S^5$ strings''},
\textsf{\doiref{10.1142/S0217732307022785}{Mod.~Phys.~Lett.~A22,~415~(2007)}},
\texttt{\arxivref{hep-th/0606214}{hep-th/0606214}}.
%
%%CITATION = HEP-TH 0604175;%%
\bibitem{Dorey:2006dq}
N.~Dorey,
\textit{``Magnon bound states and the AdS/CFT correspondence''},
\textsf{\doiref{10.1088/0305-4470/39/41/S18}{J.~Phys.~A39,~13119~(2006)}},
\texttt{\arxivref{hep-th/0604175}{hep-th/0604175}}.
%
%%CITATION = HEP-TH 0608047;%%
\bibitem{Chen:2006gq}
H.-Y.~Chen, N.~Dorey and K.~Okamura,
\textit{``On the scattering of magnon boundstates''},
\textsf{\doiref{10.1088/1126-6708/2006/11/035}{JHEP~0611,~035~(2006)}},
\texttt{\arxivref{hep-th/0608047}{hep-th/0608047}}.
%
%%CITATION = HEP-TH 0608049;%%
\bibitem{Roiban:2006gs}
R.~Roiban,
\textit{``Magnon bound-state scattering in gauge and string theory''},
\textsf{\doiref{10.1088/1126-6708/2007/04/048}{JHEP~0704,~048~(2007)}},
\texttt{\arxivref{hep-th/0608049}{hep-th/0608049}}.
%
%%CITATION = HEP-TH 0610295;%%
\bibitem{Chen:2006gp}
H.-Y.~Chen, N.~Dorey and K.~Okamura,
\textit{``The asymptotic spectrum of {$\mathcal{N}=\mathord{}$4} super
  Yang-Mills spin chain''},
\textsf{\doiref{10.1088/1126-6708/2007/03/005}{JHEP~0703,~005~(2007)}},
\texttt{\arxivref{hep-th/0610295}{hep-th/0610295}}.
%
%%CITATION = JPAGB,A25,L991;%%
\bibitem{Zhang:1992aa}
R.~B.~Zhang,
\textit{``A two-parameter quantization of osp(4$/$2)''},
\textsf{\doiref{10.1088/0305-4470/25/16/001}{J.~Phys.~A25,~L991~(1992)}}.
%
\bibitem{Yamane:1994aa}
H.~Yamane,
\textit{``Quantized enveloping algebras associated with simple Lie
  superalgebras and their universal $R$-matrices''},
\textsf{Publ.~Res.~Math.~Inst.~Sci.~1,~15~(1994)}.
%
%%CITATION = Q-ALG/9603015;%%
\bibitem{Yamane:1999aa}
H.~Yamane,
\textit{``On defining relations of affine Lie superalgebras and affine
  quantized universal enveloping superalgebras''},
\textsf{Publ.~Res.~Math.~Inst.~Sci.~3,~321~(1999)},
\texttt{\arxivref{q-alg/9603015}{q-alg/9603015}}.
%
%%CITATION = MATH.QA/0304406;%%
\bibitem{Yamane:2003aa}
H.~Yamane,
\textit{``A central extension of $U_q(sl(2|2)^{(1)})$ and $R$-matrices with a
  new parameter''},
\textsf{\doiref{10.1063/1.1616251}{J.~Math.~Phys.~11,~5450~(2003)}},
\texttt{\arxivref{math.QA/0304406}{math.QA/0304406}}.
%
%%CITATION = ARXIV:0705.1071;%%
\bibitem{Heckenberger:2007aa}
I.~Heckenberger, F.~Spill, A.~Torrielli and H.~Yamane,
\textit{``Drinfeld second realization of quantum affine superalgebras of
  $D^{(1)}(2,1;x)$ via the Weyl groupoid''},
\texttt{\arxivref{0705.1071}{arxiv:0705.1071}}.
%
%%CITATION = ARXIV:0802.0777;%%
\bibitem{Beisert:2008tw}
N.~Beisert and P.~Koroteev,
\textit{``Quantum Deformations of the One-Dimensional Hubbard Model''},
\texttt{\arxivref{0802.0777}{arxiv:0802.0777}}.
%
%%CITATION = HEP-TH/0412163;%%
\bibitem{Wolf:2004hp}
M.~Wolf,
\textit{``On hidden symmetries of a super gauge theory and twistor string
  theory''},
\textsf{\doiref{10.1088/1126-6708/2005/02/018}{JHEP~0502,~018~(2005)}},
\texttt{\arxivref{hep-th/0412163}{hep-th/0412163}}.
%
%%CITATION = HEP-TH/0608225;%%
\bibitem{Popov:2006qu}
A.~D.~Popov and M.~Wolf,
\textit{``Hidden symmetries and integrable hierarchy of the
  {$\mathcal{N}=\mathord{}$4} supersymmetric Yang-Mills equations''},
\textsf{\doiref{10.1007/s00220-007-0296-1}{Commun.~Math.~Phys.~275,~685~(2007)%
}},
\texttt{\arxivref{hep-th/0608225}{hep-th/0608225}}.
%
%%CITATION = HEP-TH 0604135;%%
\bibitem{Hofman:2006xt}
D.~M.~Hofman and J.~M.~Maldacena,
\textit{``Giant magnons''},
\textsf{\doiref{10.1088/0305-4470/39/41/S17}{J.~Phys.~A39,~13095~(2006)}},
\texttt{\arxivref{hep-th/0604135}{hep-th/0604135}}.
%
%%CITATION = HEP-TH 0612079;%%
\bibitem{Maldacena:2006rv}
J.~Maldacena and I.~Swanson,
\textit{``Connecting giant magnons to the pp-wave: An interpolating limit of
  $AdS_5\times S^5$''},
\textsf{\doiref{10.1103/PhysRevD.76.026002}{Phys.~Rev.~D76,~026002~(2007)}},
\texttt{\arxivref{hep-th/0612079}{hep-th/0612079}}.
%
\end{thebibliography}
\bibliographystyle{nb}

\end{document}